# Computational prediction and analysis of protein-protein interaction networks

by

Somaye Hashemifar

A thesis submitted in partial fulfillment of the requirements for

The degree of

Doctor of Philosophy in Computer Science

at the

TOYOTA TECHNOLOGICAL INSTITUTE AT CHICAGO

August, 2017

Thesis Committee:
Jinbo Xu (Thesis advisor)
Madhur Tulsiani,
Stefan Canzar

# Computational prediction and analysis of protein-protein interaction networks

a thesis presented

by

Somaye Hashemifar

In partial fulfillment of the requirements for the degree of
Doctor of Philosophy in Computer Science
Toyota Technological Institute at Chicago
August, 2017

-Thesis Committee-

Madhur Tulsiani 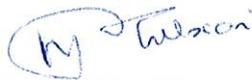
Committee member           Signature      Date

Stefan Canzar 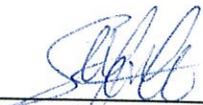
Committee member           Signature      Date

Jinbo Xu 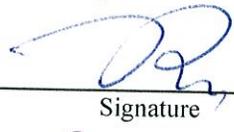
Thesis/ Research Advisor   Signature      Date

Avrim Blum 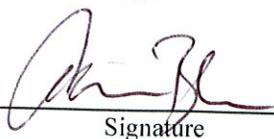
Chief Academic Officer     Signature      Date

# Computational prediction and analysis of protein-protein interaction networks

by

Somaye Hashemifar

## Abstract

Biological networks provide insight into the complex organization of biological processes in a cell at the system level. They are an effective tool for understanding the comprehensive map of functional interactions, finding the functional modules and pathways. Reconstruction and comparative analysis of these networks provide useful information to identify functional modules, prioritization of disease causing genes and also identification of drug targets. The talk will consist of two parts. I will discuss several methods for protein-protein interaction network alignment and investigate their preferences to other existing methods. Further, I briefly talk about reconstruction of protein-protein interaction networks by using deep learning.

# Acknowledgements

I thank my advisor Jinbo Xu who was always supportive and positive to help me find my own path for my thesis. I thank my committee members Madhur Tulsiani and Stefan Canzar for their support and suggestions. My sincere thanks are also to all professors from TTI-C and University of Chicago for their encouraging and sharing their knowledge on various subjects and courses.

I also thank all my dear friends for their friendship and constantly support. They made the PhD years a joyful experience for me.

Finally, my biggest thanks go to my loving and caring family, especially my dearest Behnam. I will thank you in person.

# Contents

# Chapter 1
# Introduction

It has been observed that proteins do not act alone but usually interact with one another to carry out specific biological functions. Proteins are often assembled into complexes that perform specific functions related to structure, metabolism, growth and communication. Protein-protein interactions (PPIs) occur when two proteins physically bind together to form functional modules and pathways that carry out most cellular processes. These interacting patterns form the PPI network. Figure 1.1 presents the PPI network of yeast[1]. PPI networks provide insight into the complex organization of biological processes in a cell at the system level. They are an effective tool for understanding the comprehensive map of functional interactions, and for identifying functional modules and pathways.

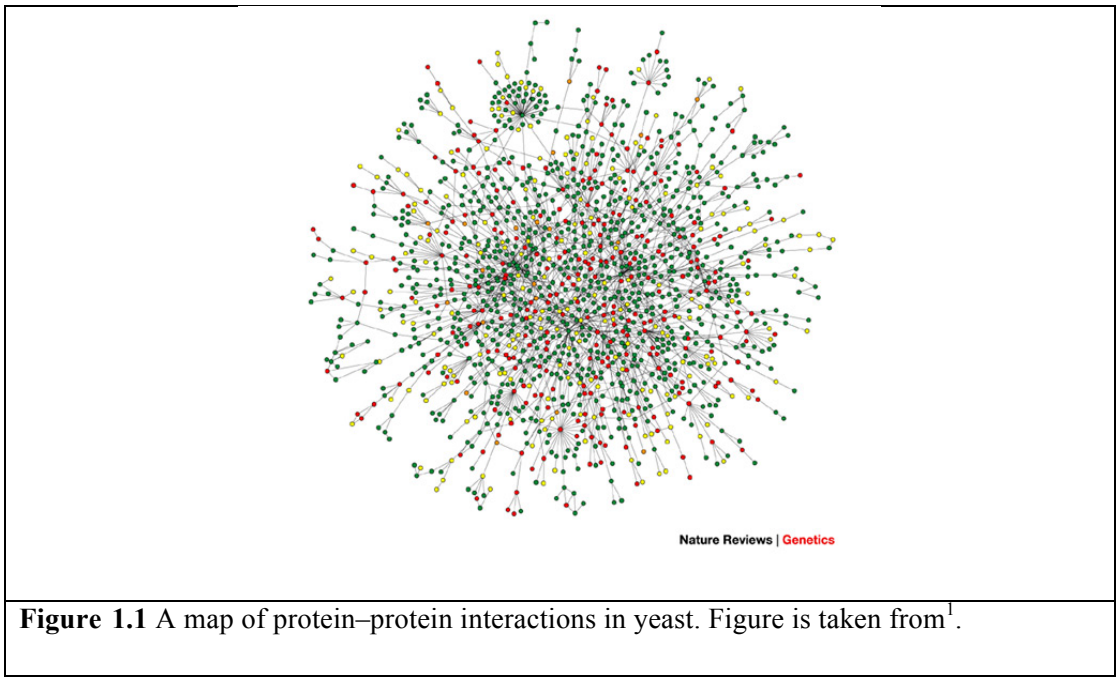

**Figure 1.1** A map of protein–protein interactions in yeast. Figure is taken from[1].

A PPI network contains some topologically and functionally important proteins such as hubs and bottlenecks. Hub proteins have many interactions, may be involved in various biological modules and play a central role in all biological processes. Bottlenecks refer to those proteins with a high betweenness centrality (i.e., the number of shortest paths passing through a node)[2]. These proteins usually connect functional clusters, so removing them can divide a PPI network into several sub-networks and disrupt the cooperation between functional modules. PPI networks are known to be hub-regularized, meaning that there are only a few hubs while other nodes have a small degree.

Inference of PPIs is essential for understanding the dynamic properties of such processes like metabolic pathways, signaling cascades, DNA transcription and replication, DNA translation and many additional processes. It also can aid significantly in identifying the function of newly discovered proteins and simplifying the discovery of new drug targets. Therefore, one of the major goals in functional genomics is to determine the complete map of interactions among proteins (i.e. interactome) that can occur in a cell.

Various experimental techniques such as yeast-two-hybrid[3] and protein co-immunoprecipitation[4] have been developed for detecting the protein interactions in different species including which could lead to the identification of the functional relationships between proteins. These methods have resulted in detecting hundreds of potential interacting proteins in several species such as Yeast [5], Drosophila [6], and Helicobacter-pylori [7]. However, experimental methods are very expensive, significantly time consuming and labor-intensive. Moreover, high throughput experimental results have shown both high false positive beside false negative interactions for protein [8]. The limitations associated with the mentioned biochemical approaches have resulted in developing many computational methods for both large-scale prediction of PPIs and validation of experimental data.

As an increasing amount of protein–protein interaction data becomes available, their computational interpretation has become an important problem in bioinformatics. The study of PPI networks, such as comparative analysis, facilitates the detection of evolutionary and functionally conserved pathways or complexes and the prediction of protein function. Network alignment provides a bridge to transfer knowledge from well-studied species such as yeast or worm to less well-studied species such as human. This is very important, because many crucial biological process and diseases in human are hard to study experimentally. Besides cross-species transfer of functional knowledge, network alignment is used to predict phylogenetic relationships of different species based on similarities between their PPI networks.

In my thesis, I study two major problems in this context, PPI network alignment and PPI network reconstruction. The Main contributions are:

a) The role of essential proteins in network alignment (chapter 3). We design experiments to highlight the role of important proteins such as hubs and bottlenecks in finding the functionally similar proteins and aligning them together.
b) The role of modularity in network alignment (chapter 4). We show how modules are useful for determining the function of less-known. This

information will enhance both the identification of interologs and scoring the reliability of the interactions in a PPI network.

c) Modeling the network alignment by convex optimization to generalize it to multiple networks (chapter 5). We describe how modeling the problem as optimization can improve the alignment of multiple PPI networks.

d) The utilization of similarities between different organs for predicting their networks (chapter 6). We combine the small data available for different organs, resulting in a larger set of data, to reconstruct a more reliable network for each one of them.

e) Applying convolutional neural networks for inference of protein-protein interaction networks (chapter 7). We present experiments to compare the ability of deep learning in network prediction compared to the other ML-based methods.

# Chapter 2

# Background and Preliminaries

## 2.1 Protein-protein interaction network

PPI networks are represented by a graph $G = (V, E)$ where $V$ is the set of vertices (proteins) and E the set of edges (interactions). Let $N(u)$ denote the neighbors of a node $u \in V$, $|N(u)|$ the size of $N(u)$ and $deg(u)$ the degree of vertex $u$, i.e., $deg(u) = |N(u)|$. Each edge $e = (u, v) \in E$ may be associated with a score indicating the interaction.

It is well known [9] that PPI networks have a hierarchical structure, which can be represented by a binary tree with leaves corresponding to the proteins and each internal non-leaf node to the clusters. Each cluster contains the proteins at the leaves of the subtree rooted at the corresponding node, split between its left and right child. There is different methods like HAC-ML [10] to infer the binary tree underlying a given network in which clusters have two main properties: (i) Proteins within a cluster are relatively more densely connected than proteins in different clusters. (ii) Every cluster, except for those close to the root, corresponds to a specific network motif with its proteins performing similar functions. Figure 2.1 presents an example of hierarchical structure of a toy PPI network.

The hierarchical structure of network $G$ is represented by a set of clusters $C = \{cl_1, cl_2, ..., cl_m\}$. Each cluster $cl$ consists of a subset of proteins in the corresponding network.

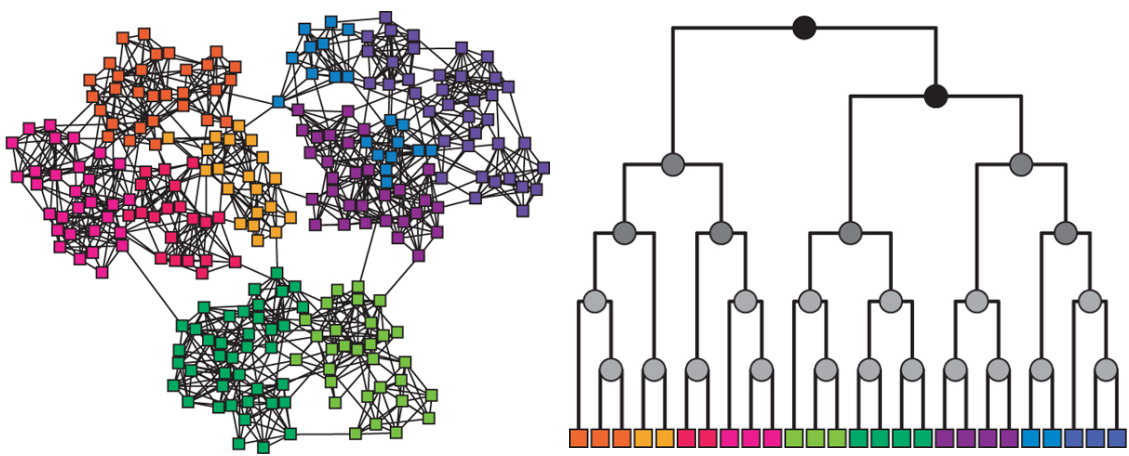

**Figure 2.1** Hierarchical structure of a sample network and its corresponding binary tree. Each internal node represents a cluster. Figure is taken from [11].

## 2.2 PPI network inference

In the advent of in vitro studies, a huge number of PPIs has been detected at the whole proteomes scale. The growing increase of the available number of known protein-protein interactions makes Machine Learning (ML) approaches the tool of choice to compute statistical significance of the predicted PPIs and estimate new protein-protein interactions. ML-based methods commonly view PPI prediction task as a binary classification problem, with each instance being a pair of proteins either interacting or not. These methods require sufficient number of training instances and features to maximize the classification accuracy of the prediction algorithms. Therefore, it is substantial to identify various protein features and to train a reliable machine learning algorithm.

A variety of protein features have been used by different studies either individually or in combination to infer PPIs. These features can be categorized in several divisions: Primary sequence-based, structure-based, protein expression-based, Functional association-based, network topology-based, evolutionary conservation-based and essentiality-based features. Primary sequence-based features, which are the most commonly used ones rely on the observation that two proteins are likely to interact with each other if they have similar sequences [12]. Structure-based characteristics take into account the known secondary or three-dimensional structures and domains of interacting proteins. Protein expression-based properties are based on the intuition that proteins with physical interaction have correlated transcription profiles. Functional association-based features take advantages of the observations that protein-pairs acting in same biological processes are very likely to interact with each other. Various methods have deployed different type of functional similarities including Gene Ontology, KEGG orthology and MIPS functional catalogue to predict novel PPIs. Network topology-based features exploit the evidence that two proteins with similar local topology are very likely to interact with each other [13]. The hypothesis behind the evolutionary conservation-based features is that two interacting proteins should have a high chance to share correlated evolutionary history. The essentiality-based features are based on the hypothesis that interacting proteins are either essential or non-essential but not both. Different ML methods choose different learning algorithms to combine multiple information source [14, 15]. These techniques include support vector machine [16-19], logistic regression, Bayesian networks [20, 21], decision tree [22, 23], random forest [24, 25], K nearest neighborhood, conditional random fields and artificial neural networks to name a few. Among these, the mostly used ones are SVMs and Bayesian networks.

## 2.3 PPI network alignment

PPI network alignment aims to find an overall match between proteins from different species by using both the sequence similarity between proteins, as well as the topology of PPI networks. Resulting mapping can be used for transferring the knowledge of protein functions from well-studied species to other ones. PPI network alignment is computationally intractable due to NP-completeness of the underlying subgraph isomorphism problem. Therefore, there is a need for proposing heuristic methods to solve this problem.

PPI networks can be aligned either locally or globally. Local network alignment (LNA) such as NetworkBlast [26] and AlignNemo [27] aims to find small isomorphic sub-networks corresponding to pathways and protein complexes and thus, may yield a many-to-many mapping between the proteins. These methods search for conserved sub-networks, in which nodes correspond to groups of orthologous proteins and edges to conserved interactions. Different from LNA, global network alignment (GNA) aims to maximize the overall match between the input networks. Such methods such as IsoRank [28], GHOST [29] and NETAL [30] are designed for pairwise alignment while others such as IsoRankN [31], SMETANA [32], NetCoffee [33], BEAMS [34] and FUSE [35] for multiple alignment. Figure 2.2 shows the global and local network alignments between a pair of toy PPI networks

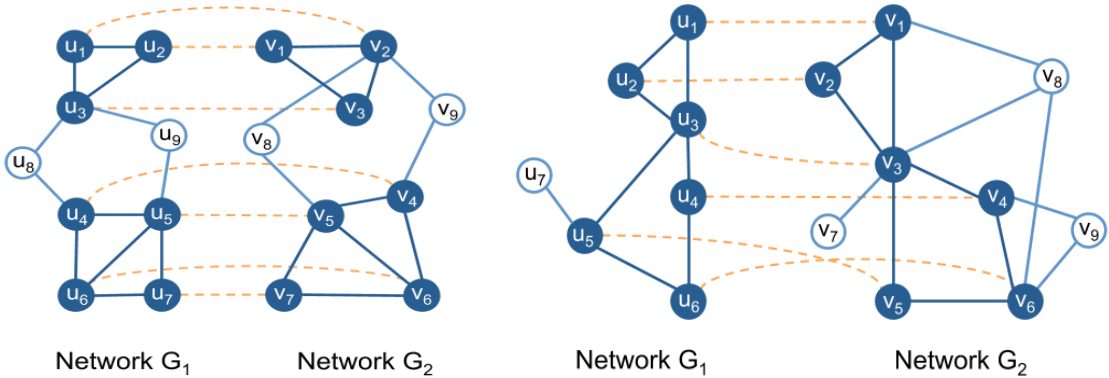

**Figure 2.2** A global network alignment (left) versus a local network alignment (right) for two sample PPI networks. Figure is taken from [36].

In this thesis, we focus on the task of finding a global alignment between either a pair of or multiple PPI networks. That is given a set of $N$ PPI networks $G_i = (V_i, E_i), 1 \leq i \leq$ $N$, the alignment problem is finding an optimal decomposition $\mathcal{A}$ of all nodes $\mathcal{V} = \cup_{i=1}^{N} V_i$ such that $\mathcal{V} = \mathcal{A}_1 \cup \mathcal{A}_2 \cup ... \cup \mathcal{A}_K$ where each $\mathcal{A}_j$ contains at most one protein from each network and any two $\mathcal{A}_j$ and $\mathcal{A}_l$ are disjoint. Each $\mathcal{A}_j$ in the alignment is called a group or a cluster. Proteins in each cluster are mutually aligned to one another. In case of aligning of two PPI networks, one may represent the alignment with a function $g = V_1 \rightarrow V_2$ that maps node set $V_1$ to $V_2$.

The quality of resulting global alignment can be evaluated by several functional consistency and topological measures [37-39]. Functional consistency metrics are particularly important in the context of network alignments since one of their main purposes is the transfer of functional annotations and modules between networks. We describe several of the most important measures that is used for evaluation so far.

We employ Gene Ontology (GO) terms and KEGG Orthology (KO) annotations to measure the functional consistency of an alignment and examine the conservation of pathways by an alignment. GO terms describe roles of proteins in terms of their associated biological process (BP), molecular function (MF) and cellular component (CC). KO annotations integrate pathway and genomic information in KEGG which is well known for its comprehensive pathway database [40]. KO represents a group of orthologous genes and its direct link to KEGG allows for the identification of pathways that might provide therapeutic targets. The functional consistency measures are based on the observation that orthologous genes are often share identical GO terms and KO annotations.

**Precision**: A cluster is annotated if at least two of its proteins have either KO annotations or GO terms assigned. An annotated cluster is consistent if all its proteins share at least one common annotation. Precision is defined as the ratio of consistent clusters among to annotated clusters.

**Recall**: Recall is defined as the total number of proteins in consistent clusters divided by the total number of proteins assigned at least one annotation.

**Average of functional similarity** (AFS): AFS is based on the semantic similarity of GO terms, which depends on their distance in the ontology. We use Schlicker's similarity based on the Resnik ontological similarity to calculate the functional similarity41. Schlicker's similarity is one of the best performing methods for computing the functional similarity between proteins. Let $s_{cat}(u, v)$ denote the GO functional similarity of proteins u and v in category cat (i.e. BP, MF or CC). AFS of an alignment $\mathcal{A}$ in category cat is defined as follows:

$$\text{AFS}_{cat}(\mathcal{A}) = \frac{1}{K}\sum_{l=1}^{K}\left(\frac{1}{|\mathcal{A}_l|}\sum_{v_i,v_j \in \mathcal{A}_l, i \neq j} s_{cat}(v_i, v_j)\right).$$

**Functional consistency** ($FC$). $FC$ is the fraction of aligned proteins sharing common GO terms. The larger the fraction, the more biologically meaningful the alignment is.

**Mean normalized entropy** (MNE): The normalized entropy of a cluster $\mathcal{A}_l$ is defined as:

$$\text{NE}(\mathcal{A}_l) = \frac{1}{\log(d)} \times \sum_{i=1}^{d} p_i \times \log(p_i),$$

where d is the number of different GO annotations in $\mathcal{A}_l$ and $p_i$ represents the fraction of proteins in $\mathcal{A}_l$ with annotation $GO_i$. A cluster with lower entropy is more functionally coherent. MNE of alignment $\mathcal{A}$ is the mean of normalized entropy over all annotated clusters.

**Conserved orthologous interactions($COI$)**: It is calculated as the total number of interactions between all consistent clusters. $COI$ may be a better measure than CI because it detects whether the conserved interactions are spurious or correspond to real conserved interactions be- tween orthologous proteins. An alignment with larger COI may lead to identifying functionally conserved sub-netwroks (i.e clusters) composed of orthologous genes.

We also evaluate the topological quality of an alignment by the following measures.

**Conserved Interaction (CI)**: This measure is also called Edge correctness (EC). It is calculated as the ratio of the number of aligned interactions to the number of interactions between output cluster.

**c-coverage**: It is the number of clusters composed of proteins from exactly c species. Specifically, total coverage is the number of clusters composed of proteins from at least two species. Clusters with large c explain a larger amount of data better than clusters with small c.

**Symmetric substructure score** ($S^3$): The intuition underlying $S^3$ is to penalize the alignments that map sparse regions of the network to denser ones and vise-versa 39. Let G[V] denote the induced sub-network of G with node set V and E(G) denote the edge set of network G. Let $f(E_1) = \{(g(u), g(v)) \in E_2 : (u, v) \in E_1\}$ and $f(V_1) = \{g(v) \in V_2 : v \in V_1\}$. Mathematically, $S^3$ is defined as follows.

$$S^3 = \frac{|f(E_1)|}{|E_1| + |E(G_2[f(V_1)])| + |f(E_1)|} \times 100$$

**Largest common connected subgraph** (LCCS): It is calculated as the number of edges in the largest connected subgraph in an alignment. Larger and denser subgraphs give more insight into common topology of the network. In addition, the larger and denser subgraphs may be more biologically important, as Bader has shown that a dense PPI sub-network may correspond to a vital protein complex [42].

Among the above measures, $S^3$ and LCCS have been used only for evaluation of pairwise alignments.

# Chapter 3

# Global Alignment of PPI Networks based on their essential proteins

## 3.1. Introduction

A biological network usually contains some topologically and functionally important proteins such as hubs and bottlenecks. Hub proteins have many connections, may be involved in various biological modules and play a central role in all biological processes. In Han's work [43], proteins with more than five interactions are defined as hubs, while those with fewer interactions are peripheral nodes. Bottlenecks refer to those proteins with a high betweenness centrality (i.e., the number of shortest paths passing through a node) [2]. These proteins usually connect functional clusters, so removing them can divide a PPI network into several sub-networks and disrupt the cooperation between functional modules [44]. Since hubs and bottlenecks are topologically and functionally important, they tend to mutate more slowly and thus, are more conserved. That is, they are more likely to be aligned. To make use of this observation, we assign a score or weight to each node and edge of a PPI network using an iterative minimum-degree heuristics algorithm, measuring the topological and functional importance of a node (i.e., the likelihood of being a hub or bottleneck) and an edge in the PPI network with respect to the global network topology. Such an importance score reflects the global topological property of a protein. Then we calculate an alignment score for a pair of proteins using two properties: their relative importance scores (i.e., global topological property) and sequence information. Meanwhile, the global topological property is the most important and informative. Finally, we construct a global network alignment by picking those protein pairs with high alignment scores using a greedy method.

## 3.2. Method

### 3.2.1. Computing the topological and functional importance of proteins

We calculate the relative importance of a node or edge based upon only the network topology information of a PPI network. Such a relative importance shows the role of a node or edge in maintaining network structure or function[45]. Although high-degree nodes play an important role in maintaining the structure and function of a network [46], we do not simply use the degree of one node to calculate its relative importance since the degree is only a local property. We want a global topological property reflecting the structure of the entire network.

We do not use existing measures such as edge-betweenness [47] either, which defines the number of the shortest paths going through an edge in a network. That is, edge-betweenness takes into consideration only the shortest paths in a graph. Nevertheless, for the robustness of a network the longer alternative paths are also important [47]. In addition, it is also observed that 1) edges connecting high-degree nodes are more important since they connect many nodes and may be relevant to the global structure property of the network [48]. 2) a pair of two nodes with a large number of common neighbors are more likely to be related [49].

Here we use a minimum-degree heuristics algorithm to calculate the topological importance of nodes and edges, starting from the nodes with degree one and stopping at those with degree d. The value of d cannot be very large since the deletion of very high-degree nodes (e.g., hubs) may destroy the whole network functionally or structurally while random deletion of a fraction of peripheral nodes may cause only a small damage to the network [45, 50]. Empirically $d = 10$ yields a good result. To calculate the relative importance of nodes, we assign an initial weight to nodes and edges as follows.

$$w(e) = \begin{cases} 1 & e \in E \\ 0 & otherwise \end{cases}, \quad w(u) = 0 \quad \forall u \in V$$

where $w(e)$ and $w(u)$ represent the weight of edge $e$ and node $u$, respectively. We may initialize the edge weight by the PPI confidence score if it is available in the PPI data.

We update the weight by always removing one of the nodes with minimum degree. When one node is removed, its adjacent edges are also removed and the weight of the removed node and edges are allocated to their neighboring nodes and edges. In this way, the topological information is propagated from a node to its neighbors. In particular, when removing node $u \in V$, we update the weights as follows.

$$\begin{cases} \forall v \in N(u): w(v) = w(v) + w(u) + w(u,v) & \deg(u) = 1 \\ \forall v_1, v_2 \in N(u): w(v_1, v_2) = w(v_1, v_2) + \dfrac{w(u) + \sum_{v \in N(u)} w(u,v)}{\dfrac{|N(u)||N(u)-1|}{2}} & \deg(u) > 1 \end{cases}$$

Figure 3.1 shows for a small example PPI network how an edge gains more weight after the removal of some peripheral nodes. For example, when nodes d, c, e and f are removed, their own weight and those of their adjacent edges are transferred to the edge (a, b), which indicates that this edge is important in maintaining the network connectivity. After calculating the weights, we assign an importance score as follows to each node by combining both node and edge weight to indicate its topological importance in the network.

$$S(v) = w(v) + \lambda \sum_{u \in V} w(u,v),$$

where $S(v)$ is the score of node $v$, $\lambda$ controls the importance of the edge weight relative to the node weight. Empirically $\lambda = 0.1$ yields a biologically more meaningful alignment. Finally, we normalize $S(v)$ as follows to reduce the impact of network size.

$$S(v) = \frac{S(v)}{max_{v \in V}\{S(v)\}}.$$

The way we calculate the relative importance of nodes and edges is inspired by graph tree-decomposition, which is used to simplify a graph as a tree in which each vertex represents a highly-connected subgraph component and each edge represents the intersection between two adjacent components. The size of the highly-connected components reflects the topological complexity of a graph and also importance of nodes. Several simple heuristics methods such as the minimum-degree heuristic method [51, 52] are developed to tree-decompose a general graph.

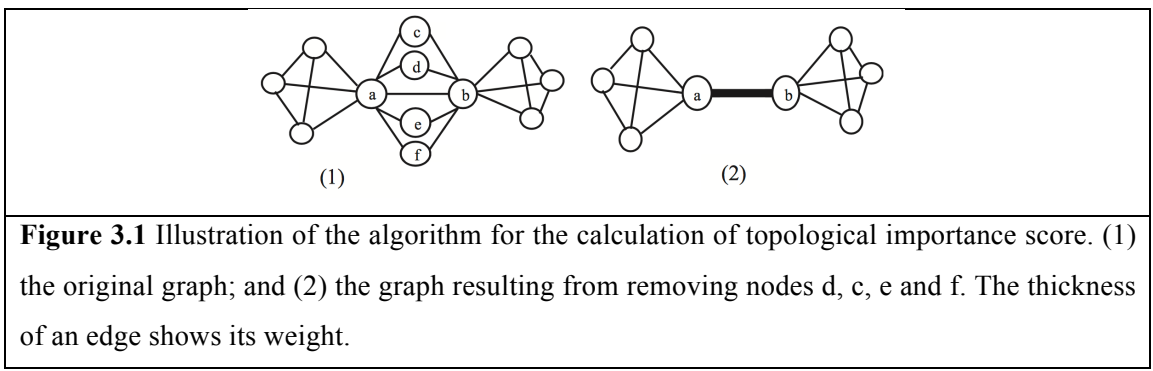

**Figure 3.1** Illustration of the algorithm for the calculation of topological importance score. (1) the original graph; and (2) the graph resulting from removing nodes d, c, e and f. The thickness of an edge shows its weight.

**Remark.** To validate that the resultant importance score (i.e., $S$) makes biological sense, we examine the top 50 proteins with the highest S scores in the human PPI network. Meanwhile, all the top 10 proteins have a very high degree, which indicates they are vital hubs of the network. The two example proteins are P62993 with degree 663 and Q9H0R8 with degree 491. See Figure 3.2(a) for the sub-network containing P62993. On the other hand, among the top 50 proteins there are also some low-degree proteins, such as Q9UPN3 with degree 7. As shown in Figure 3.2(c), although Q9UPN3 is not a hub, it is a bottleneck connecting several functional modules. This protein is also related to breast cancer disease [53]. Another interesting example is P04156 with degree 52 and betweenness 0.005. As shown in Figure 3.2(b), this protein is a hub connecting several other hubs. Removal of this protein can disrupt the cooperation of the hubs connecting to it.

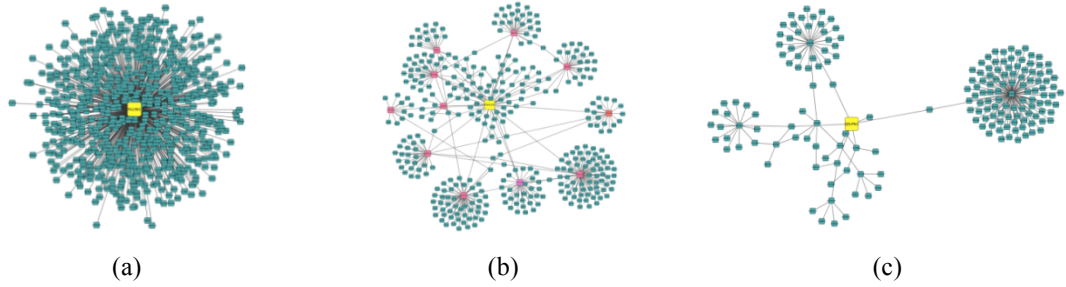

(a) (b) (c)

**Figure 3.2** Three example proteins (in yellow) with high importance scores in the human PPI network. (a) Protein P62993, which has the largest degree; (b) Protein P04156, which connects to some hubs (in red); (c) Protein Q9UPN3 with low degree that performs as a bottleneck.

### 3.2.2. Building the alignment

The normalized $S$ score measures the relative importance of one protein with respect to the whole PPI network. It reflects the global topological properties of one protein in a network. If two nodes have similar $S$ scores, they may be similarly important in their respective networks. Thus, they are more likely to be aligned. We calculate the topological similarity between two nodes $u \in V_1$ and $v \in V_2$ as follows:

$$A(u, v) = min(S(u), S(v)).$$

We also incorporate sequence homology information (i.e., sequence similarity) into our alignment score. Let $B(u, v)$ denotes the normalized BLAST bitscore for two proteins $u$ and $v$. The final alignment score is defined as follows.

$$A^*(u, v) = \alpha \times A(u, v) + (1 - \alpha) \times B(u, v),$$

where $0 \leq \alpha \leq 1$ is a parameter that controls the contribution of sequence similarity relative to topological similarity. Meanwhile, $\alpha = 1$ implies that only topological information is used, while $\alpha = 0$ implies that only sequence information is used. Tuning α allows us to find the relative importance of sequence information in aligning the networks. In our implementation, we set α to 0.7 by default. That is, our method uses much more network topology information than sequence information.

Our algorithm identifies the pair of nodes with the highest alignment score as a seed alignment and gradually extends it using a greedy algorithm. After aligning a pair of nodes u and v, we then consider aligning their neighbors, which is reasonable because functional modules and protein complexes are densely connected and tend to be separated from other sub-network modules. Algorithm continues to align neighboring nodes until their alignment score is relatively high (more than the average of the alignment scores). When the sub-network alignment resulting from the initial seed is terminated, the next best unaligned pair is chosen as a new seed. This procedure is repeated until all proteins of the smaller network are aligned with the proteins of the larger network.

### 3.2.3. Time Complexity

Let $n = max\{|V_1|, |V_2|\}$. At the first step, it takes $O(n)$ to find the node with minimum degree. As we mentioned before, we only remove the nodes with degree less than 10. Thus, updating the weight of the neighbors can be done in $O(1)$. Further, as we can remove up to $n$ nodes from a network, the total time complexity for the first step is $O(n^2)$. At the second step, we calculate the alignment score for each pair of nodes of the input networks. Since there are at most $n^2$ pair nodes, this step takes $O(n^2)$. At the final step, a seed can be selected in $O(n^2)$. Then for extension, we use a priority queue to save the neighbors of each pair of aligned nodes. Since each node of the graph has at most $n$ neighbors, updating the priority queue takes $O(nlog(n))$. Extracting the pair with highest score from this queue can be done in constant time. That is, the final step for aligning $n$ nodes takes $O(n^2 log(n))$. As such, the total time complexity is $O(n^2 log(n))$.

### 3.3. Results

We compare our algorithm HubAlign with several popular and publicly available global network alignment methods IsoRank[54, 55], MI-GRAAL[56], NETAL [30], GHOST [29] and PISwap [57]. Following [57], we use the alignment produced by GRAAL and IsoRank as input of PISwap. We ran IsoRank and PISwap with the default parameters. MI-GRAAl was run using the degree, signature similarity and sequence similarity. The parameters for GHOST are automatically determined or set to default.

### 3.3.1. Alignment of the yeast and human PPI networks

We apply HubAlign to align the yeast and human PPI networks, which are taken from IntAct [58]. The yeast PPI network has 5673 nodes and 49830 edges and the human network consists of 9002 nodes and 34935 edges. As shown in Table 3.1, HubAlign produces an alignment with much larger EC, LCCS and $S^3$ than the other methods except NETAL. To measure the functional consistency ($FC$) and Average of functional similarity ($AFS$) of an alignment, we extract the GO annotations for all the involved proteins from the Gene Ontology database[59]. Some proteins may not have any GO annotations, so we just take into consideration the aligned pairs in which both proteins have GO annotations. Table 3.1 show that HubAlign yields alignments with significantly higher AFS than the other methods, especially when biological process (BP) and molecular function (MF) are considered. We also calculate the percentage of aligned pairs in which two proteins share at least one, two, three, four and five GO terms, respectively.

**Table 3.1** The EC, LCCS and $S^3$ of the human-yeast alignments generated by 6 methods.

| method | EC | LCCS | $S^3$ | AFS(BP) | AFS(MF) | AFS(CC) |
|---|---|---|---|---|---|---|
| IsoRank | 2.12 | 44 | 1.23 | 0.76 | 0.63 | 0.77 |
| MIGRAAL | 13.87 | 4832 | 8.12 | 0.63 | 0.52 | 0.72 |
| GHOST | 17.04 | 7000 | 13.59 | 0.82 | 0.66 | 0.83 |
| PISwap | 2.16 | 62 | 1.23 | 0.77 | 0.63 | 0.77 |
| NETAL | 28.65 | 9695 | 20.16 | 0.58 | 0.46 | 0.71 |
| HubAlign | 21.59 | 7240 | 14.67 | 0.95 | 0.81 | 0.88 |

Table 3.2 presents that HubAlign greatly outperforms the others in terms of $FC$. The advantage of HubAlign becomes larger when more shared GO terms are required to determine $FC$. NETAL yields more aligned proteins and interactions, but many aligned proteins are not functionally similar.

**Table 3.2** Functional consistency of the yeast-human alignments generated by HubAlign and the others.

| #shared GO terms | IsoRank | MIGRAAL | GHOST | PISwap | NETAL | HubAlign |
|---|---|---|---|---|---|---|
| ≥ 1 | 33.98 | 29.02 | 35.42 | 34.02 | 26.03 | 47.56 |
| ≥ 2 | 15.02 | 7.02 | 15.74 | 14.84 | 2.95 | 28.23 |
| ≥ 3 | 8.73 | 2.81 | 8.69 | 8.65 | 0.67 | 17.41 |
| ≥ 4 | 4.49 | 1.06 | 4.04 | 4.46 | 0.24 | 9.52 |
| ≥ 5 | 1.97 | 0.26 | 1.77 | 2.00 | 0.14 | 4.7 |

### 3.3.2. Alignment among other species

We also apply HubAlign to align PPI networks of Homo sapiens (human), Saccharomyces cerevisiae (yeast), Drosophila melanogaster (fly), Caenorhabditis elegans (worm) and Mus musculus (mouse). All these networks are obtained from IntAct [58]. Table 3.3 shows that the alignments produced by HubAlign outperform those by the other methods in term of AFS under all three categories BP, MF and CC. These results indicate that HubAlign can align more functionally similar proteins and find larger complexes that are significant either topologically or biologically.

### 3.3.3. Alignment of bacterial PPI networks

We use HubAlign to align the PPI networks of two bacterial species Campylobacter jejuni and Escherichia coli, which have the most complete PPI networks among all bacteria. The PPI network for Bacterium Campylobacter jejuni has 1111 nodes and 2988 edges [60]. Escherichia coli is a model organism for studying the fundamental cellular processes such as gene expression and signaling. The Escherichia coli PPI network has 1941 nodes and 3989 edges [61]. As shown in Table 3.4, HubAlign produces an alignment with larger $EC$, $LCCS$ and $S^3$ than the other methods except NETAL. In terms of $AFS$, HubAlign outperforms the other methods although all the AFS values are small due to insufficient GO annotations of the bacterial proteins. The average number of GO terms associated with the proteins of Escherichia coli and Campylobacter jejuni is much smaller than that of the other species.

### 3.3.4. Running time

HubAlign is much more computationally efficient than the others. Tested on the yeast-human alignment on a 1400 MHz Linux system with 2GB RAM, it takes NETAL, HubAlign, IsoRank, MI-GRAAL and GHOST 80, 412, 7610, 78525 and 3037 seconds, respectively, to terminate. PISwap has almost the same running time as IsoRank because the former only slightly refines the result generated by the latter.

| Table 3.3 Performance of HubAlign and the other methods in terms of AFS | | | | | | | |
|---|---|---|---|---|---|---|---|
| Alignment | AFS | IsoRank | MIGRAAL | GHOST | PISwap | NETAL | HubAlign |
| human-mouse | BP | 1.32 | 0.84 | 1.58 | 1.32 | 0.73 | **2.02** |
| | MF | 1.23 | 0.84 | 1.50 | 1.23 | 0.70 | **1.74** |
| | CC | 1.08 | 0.76 | 1.20 | 1.08 | 0.66 | **1.49** |
| mouse-fly | BP | 0.73 | 0.62 | 0.84 | 0.73 | 0.50 | **1.07** |
| | MF | 0.61 | 0.50 | 0.75 | 0.61 | 0.33 | **0.97** |
| | CC | 0.53 | 0.42 | 0.54 | 0.53 | 0.34 | **0.72** |
| mouse-yeast | BP | 0.71 | 0.60 | 0.85 | 0.70 | 0.47 | **0.96** |
| | MF | 0.64 | 0.54 | 0.80 | 0.64 | 0.36 | **0.91** |
| | CC | 0.77 | 0.67 | 0.84 | 0.40 | 0.57 | **0.91** |
| fly-yeast | BP | 0.48 | 0 | 0.54 | 0.48 | 0.38 | **0.68** |
| | MF | 0.35 | 0 | 0.42 | 0.35 | 0.23 | **0.58** |
| | CC | 0.40 | 0 | 0.44 | 0.40 | 0.36 | **0.50** |
| human-fly | BP | 0.53 | 0 | 0.61 | 0.53 | 0.41 | **0.72** |
| | MF | 0.43 | 0 | 0.54 | 0.43 | 0.28 | **0.65** |
| | CC | 0.38 | 0 | 0.41 | 0.37 | 0.30 | **0.48** |
| mouse-worm | BP | 0.63 | 0.50 | 0.67 | 0.63 | 0.42 | **0.76** |
| | MF | 0.64 | 0.46 | 0.67 | 0.64 | 0.31 | **0.81** |
| | CC | 0.40 | 0.31 | 0.41 | 0.40 | 0.25 | **0.49** |
| human-worm | BP | 0.52 | 0.43 | 0.60 | 0.52 | 0.40 | **0.64** |
| | MF | 0.34 | 0.25 | 0.40 | 0.34 | 0.23 | **0.70** |
| | CC | 0.34 | 0.27 | 0.40 | 0.34 | 0.25 | **0.44** |
| worm-fly | BP | 0.51 | 0.34 | 0.55 | 0.50 | 0.31 | **0.57** |
| | MF | 0.48 | 0.22 | 0.52 | 0.47 | 0.18 | **0.54** |

|  | CC | 0.26 | 0.14 | 0.28 | 0.25 | 0.13 | **0.31** |
|---|---|---|---|---|---|---|---|
|  | BP | 0.38 | 0.31 | 0.41 | 0.37 | 0.26 | **0.43** |
| worm-yeast | MF | 0.34 | 0.25 | 0.40 | 0.34 | 0.23 | **0.41** |
|  | CC | 0.30 | 0.25 | 0.31 | 0.30 | 0.24 | **0.32** |

**Table 3.4** The $EC$, $LCCS$ and $AFS$ of the alignments by different algorithms for the bacterial PPI networks.

| Algorithm | | | | | | |
|---|---|---|---|---|---|---|
| IsoRank | 8.50 | 11 | 1.51 | 0.20 | 0.16 | 0.07 |
| MI-GRAAL | 23.86 | 400 | 15.89 | 0.14 | 0.12 | 0.04 |
| GHOST | 23.86 | 440 | 15.03 | 0.19 | 0.14 | 0.06 |
| PISwap | 17.87 | 289 | 1.83 | 0.11 | 0.08 | 0.02 |
| NETAL | 32.36 | 661 | 19.54 | 0.10 | 0.07 | 0.02 |
| HubAlign | 24.56 | 474 | 16.51 | 0.25 | 0.22 | 0.07 |

## 3.4. Evaluation of parameter $\lambda$ and $\alpha$

Our algorithm makes use of two parameters λ and α. λ determines the relative importance of edge and node weight, while α determines the relative importance of sequence and topological similarity. In this section, we study the relationship between these two parameters and network alignment quality. We apply HubAlign to PPI networks of yeast and human and report $EC$, $LCCS$, $S^3$ and $AFS$ of their alignment for different values of parameter λ between 0 and 1. Figure 3.3 indicates that $AFS$ increases as λ gets close to 1. The underlying reason could be that the higher values of λ give more importance to the edge weights which in turn, makes the proteins with important interactions align together. On the other hand, by increasing the value of λ, we put less emphasis on the node weight and therefore, it is less likely that the hubs be aligned together. Therefore, the topological qualities (i.e. $EC$, $LCCS$ and $S^3$) decrease. Figure 3.3 (left) shows that increasing λ from 0 to 0.2 improves the $AFS$ significantly but does not change the $EC$ much. However, as we continue to increase λ further, the $EC$ decreases sharply. We also observe a slight increase in the biological quality. Thus, we can achieve a good tradeoff between the topological and the biological quality by setting λ in the range (0.1, 0.2).

Furthermore, we compute the yeast-human alignment for different values of α. As shown in Fig. 4, increasing α from 0 to 1 decreases $AFS$. This is because a larger value of α reduces the effect of sequence information. Moreover, in line with our expectations as α goes up, so does the topological quality of the alignment. Figure 3.3 (right) shows that increasing α from 0 to 0.7 does not change $AFS$ much but improves the $EC$ significantly. However, as we continue to increase α further, the $AFS$ decreases sharply. Thus, we can achieve a good tradeoff between the topological and the biological quality by setting α in the range (0.7, 1).

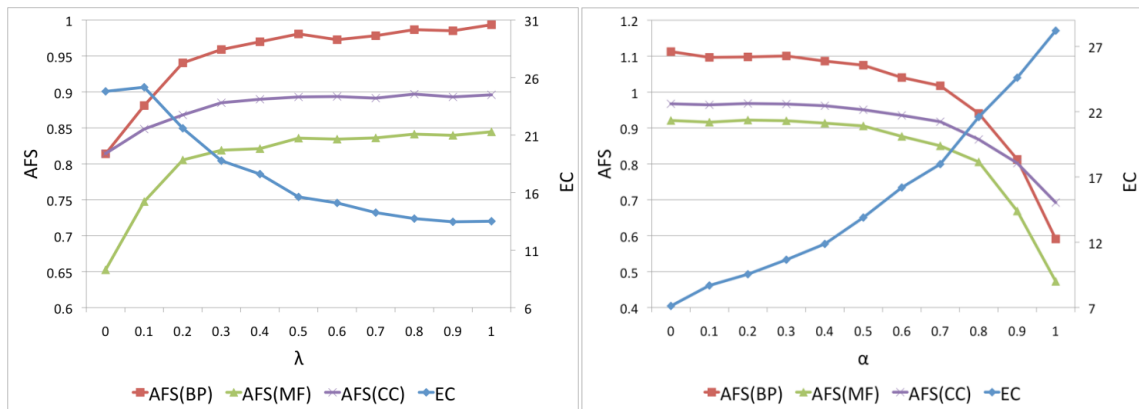

**Figure 3.3** Performance of HubAlign in terms of $AFS$ and $EC$ with respect to λ and $\alpha$. Each curve consists of 11 points corresponding to 11 different values: 0, 0.1, …, 1 from top to bottom.

# Chapter 4

# Module-based Global Alignment of PPI Networks

## 4.1. Introduction

Proteins with similar sequences are more likely to have a similar function [62]. Most of the network alignment methods thus consider sequence similarity (e.g. blast bit score) to detect homology between proteins. However, this score is not sensitive enough for remote homologs such that they may miss functionally similar proteins. In addition to the sequence of a protein, its interaction partners within a module or cluster indicate its function. Therefore, the homology of the partners of a gene in its corresponding modules can help to assess the homology of two proteins. Based on this observation, we compute a novel homology score that takes into account the local neighborhood within hierarchical clusters of a PPI network. This score is based on the observation that proteins with similar functions tend to form densely-connected sub-networks [10]. We combine the homology score and a previously introduced topology score[38] into an alignment score and propose a novel iterative strategy to optimally match proteins. Starting from an optimal bipartite matching of proteins, ModuleAlign iteratively selects highest-scoring protein pairs and adjusts alignment scores in their neighborhood to promote conservation of interactions. This work introduces ModuleAlign, a novel pairwise global network alignment approach. Its novel scoring scheme integrates sequence information and both local and global network topology. Based on a hierarchical clustering of the input networks, we compute a homology score between proteins. We propose an iterative approach to find an alignment that scores high in our model while trying to preserve interactions.

## 4.2. Method

### 4.2.1. Novel cluster-based homology score

We first employ HAC-ML [10] to determine the clusters of both input networks. Afterwards, we calculate the similarity between each pair of clusters $cl \epsilon C_1$ and $cl' \epsilon C_2$, where $C_1$ and $C_2$ represent the set of clusters for two input networks $G_1$ and $G_2$, respectively. Knowing that proteins within each cluster have similar functions and thus their sequences are expected to show a certain degree of similarity, we compute the similarity $clusterSim(cl, cl')$ between a pair of clusters as the average blast score between contained proteins:

$$clusterSim(cl, cl') = \frac{\sum_{u \epsilon cl, v \epsilon cl'} blast(u, v)}{|cl| * |cl'|},$$

where $blast(u, v)$ is the blast score between proteins u and v. We remove very high level clusters (i.e. those on level higher than 3) as they contain many genes. Note that *clusterSim* measures sequence similarity of proteins within functional modules of the network (see [10]). This strategy implicitly takes into account network structure and avoids the blurring of the homology signal by comparing unrelated (or weakly related) proteins. At the same time, clusters that are similar according to clusterSim contain proteins with both similar sequences (definition of *clusterSim*) and similar interaction neighborhoods (definition of clusters by [10]).

We use the hierarchical clusters to define a new homology score between proteins that does not solely rely on sequence information. We define the homology score between proteins $u$ and $v$ based on the similarity (as measured by *clusterSim*) between all clusters that contain $u$ and $v$:

$$HS(u, v) = \frac{\sum_{\substack{cl \in C_1 : u \in cl \\ cl' \in C_2 : v \in cl'}} clusterSim(cl, cl')}{|C_1| \times |C_2|}$$

Intuitively, two proteins that belong to many similar clusters are expected to have a similar function.

### 4.2.2. Novel alignment strategy

We propose an alignment score between proteins that combines our homology score with a score based on global topological similarity. The topology scores are calculated in the same way as in HubAlign [38]. For each node and every edge in the network, HubAlign calculates weights that it uses to infer the topological importance $S(v)$ for all nodes $v \in V$. Topological scores are normalized by $max_{v \in V}\{S(v)\}$. Finally, HubAlign defines the topological similarity score $TS(u,v)$ between two proteins $u \in V_1$ and $v \in V_2$ as the minimum of $S(u)$ and $S(v)$. We define the alignment score as follows.

$$A(u,v) = \alpha \times HS(u,v) + (1-\alpha)TS(u,v),$$

where $0 \leq \alpha \leq 1$ is a tradeoff parameter that controls the contribution of global topological similarity relative to the homology score. In our implementation, we empirically set $\alpha$ to 0.4 by default.

Our alignment strategy relies on a two-step process (see Figure 4.1): In the first step, we apply the Hungarian method to compute an optimal matching $M_0$ of proteins in the two networks with respect to alignment scores. The Hungarian method is a primal-dual algorithm that starts with an empty matching and iteratively increases the size of the matching using maximum-weight augmenting paths. We refer the interested to read [63] for details on the Hungarian algorithm. The goal of the first step is to align proteins that are topologically and functionally consistent. In the second step, we additionally try to maximize the number of evolutionary conserved interactions using the initial alignment as our guidance. Starting from $M_0$, we iteratively fix the heaviest edge $(u_b, v_b)$ in the current alignment, i.e. the pair of proteins with largest alignment score. For each neighbor $u$ of $u_b$, we remove $(u, g(u))$ from the current alignment and upweight the alignment score between u and all neighbors $v$ of $v_b$:

$$\forall_{v \in N(v_b)}\ A(u,v) = A(u,v) + \frac{1}{max_{u' \in V_1 \cup V_2}\{S(u')\}}$$

where $max_{u' \in V_1 \cup V_2} \{S(u')\}$ is the same normalization factor as used in the topology score. Intuitively, we increase the alignment score by one normalized unit since aligning neighbors $u$ of $u_b$ and $v$ of $v_b$ would yield one additional conserved interaction. Then, one primal-dual iteration of the Hungarian algorithm is performed to re-optimize w.r.t the updated alignment scores. This procedure is repeated until all proteins of the smaller network are matched by the final alignment $M^*$. See Algorithm 1 for more details. Without loss of generality, we assume $|V_1| < |V_2|$.

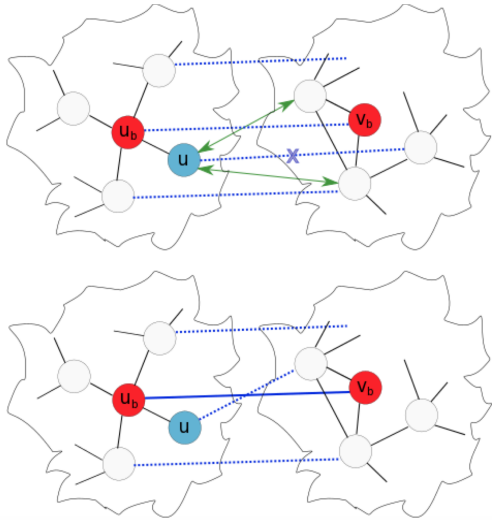

**Figure 4.1** The alignment strategy. Blue dotted lines represent the current alignment. Top: First, a pair of protein with largest score, shown in red, is selected. One of the neighbors of $u_b$, shown in blue, is selected and its mapping is removed from the alignment. Then the alignment scores between $u$ and all neighbors of $v_b$ are updated (green arrows). Bottom: Fix $(u_b, v_b)$ and run one primal-dual iteration of the Hungarian method to update the alignment (see main text for details).

### 4.2.3. Running time

Let $n = max(|V_1|, |V_2|)$. It takes $O(n^3 log(n))$ to compute the hierarchical clustering of the input networks. Moreover, calculating the topological scores takes $O(n^2)$. Thus, the time complexity for finding the alignment scores is $O(n^3 log(n))$. The first step of the alignment strategy takes $O(n^3)$. In the second step, finding the best scoring pair takes $O(n)$ and updating the alignment scores can be done in $O(n^2)$. Finally, running one iteration of the Hungarian method takes $O(n)$. Therefore, the time complexity of Algorithm 1 is $O(n^3)$, yielding an overall time complexity of ModuleAlign of $O(n^3 log(n))$.

---

Algorithm 1

---

Input: $G_1, G_2$, alignment scores $A(u,v), \forall u \in V_1, v \in V_2$

$n = |V_1|, t = 0, M^* = \emptyset$

**Step 1.**

Run the Hungarian algorithm to obtain $M^0$.

**Step 2.**

While $|M^*| < n$

1. Find $(u_b, v_b) = argmax_{(u,v)}\{A(u,v): (u,v) \in M^t\}$
2. $M^* = M^* \cup (u_b, v_b)$
3. $M' = M^t$
4. For each $u \in N(u_b)$
   i. $M' = M' \setminus \{u, g(u)\}$
   ii. $\forall v \in N(v_b): A(u,v) = A(u,v) + \frac{1}{max_{u' \in V_1 \cup V_2}\{S(u')\}}$
   iii. Starting from $M'$, fix $(u_b, v_b)$ and perform one primal-dual iteration of the Hungarian algorithm with respect to updated alignment scores (see step 2) to obtain $M''$.
   iv. $M' = M''$
5. $M^{t+1} = M'$
6. $t = t + 1$

Return $M^*$.

---

## 4.3. Result

We compare ModuleAlign with several popular and publically available global network alignment methods NETAL[30], GHOST[29], HubAlign[38], MAGNA++[64], and L-GRAAL[65]. These methods have been shown to outperform other methods such as IsoRank and MI-GRAAL on several datasets [38, 66]. As recommended by the authors, we configured the genetic algorithm implemented in MAGNA++ to optimize the $S^3$ score, running over 15000 generations with a population size of 2000. Parameters of other methods are set to their default values. We evaluate the network alignment quality by several functional consistency and topological measures proposed in different studies [37-39]. Functional consistency metrics are particularly important in the context of network alignments since one of their main purposes is the transfer of functional annotations and modules between networks.

### 4.3.1. Alignment quality

The test data includes five PPI networks for H. sapiens (human), S. cerebvisiae (yeast), D. melanogaster (fly), C. elegance (worm) and M. musculus (mouse) obtained from HINT[67]. HINT integrates the interactions from several databases, including BioGRID[68], IntAct[69], and MINT[70], and manually removes erroneous interactions. The sizes of these networks are shown in Table 4.1.

| Table 4.1 Size of the test networks | | |
|---|---|---|
| Species | Number of proteins | Number of interactions |
| human | 9336 | 29617 |
| yeast | 5169 | 20176 |
| fly | 7493 | 25674 |
| worm | 4494 | 11292 |
| mouse | 1298 | 2749 |

Figure 4.2 shows precision and recall of the alignments generated by the different methods. ModuleAlign significantly outperforms all other methods and predicts consistent classes with higher accuracy.

The consistent classes provide valuable information concerning the orthologous relationship of proteins from the two species. Taking this a step further, consistent classes allow the detection of evolutionary pathways conserved between species. In the yeast-human alignment produced by ModuleAlign, we find many conserved pathways such as RNA transport and RNA degradation. Tables 4.2 and 4.3 show the consistent classes whose KO is linked to the RNA degradation (i.e. "hsa03018" in human and "sce03018" in yeast and to the RNA transport pathway (i.e. "hsa03013" in human and "sce03013" in yeast), respectively. The definitions for all KO can be found at [71].

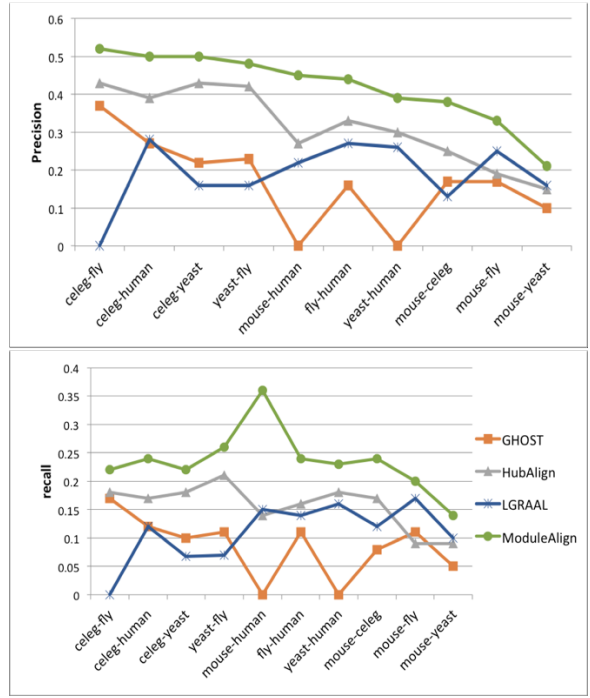

**Figure 4.2** Precision and recall on all instances. Precision and recall of NETAL and MAGNA++ is 0 (not shown).

| Table 4.2. Consistent classes in the yeast-human alignment that identify the RNA degradation ||||
|---|---|---|---|
| Yeast | Human | KO ID | KO title |
| CDC39 | CNOT1 | K12604 | CCR4-NOT transcription complex subunit 1 |
| POP2 | CNOT8 | K12581 | CCR4-NOT transcription complex subunit 7/8 |
| SSC1 | HSPA9 | K04043 | molecular chaperone DnaK |
| XRN1 | XRN1 | K12618 | 5'-3' exoribonuclease 1 |
| DIS3 | DIS3 | K12585 | exosome complex exonuclease DIS3/RRP44 |
| CDC36 | CNOT2 | K12605 | CCR4-NOT transcription complex subunit 2 |
| RRP46 | EXOSC5 | K12590 | exosome complex component RRP46 |
| RRP45 | EXOSC9 | K03678 | exosome complex component RRP45 |

| RAT1 | XRN2 | K12619 | 5'-3' exoribonuclease 2 |

**Table 4.3** Consistent classes in the yeast-human alignment that identify the RNA transport pathway.

| Yeast | Human | KO ID | KO title |
|---|---|---|---|
| SUB2 | DDX39B | K12812 | ATP-dependent RNA helicase UAP56/SUB2 |
| THO2 | THOC2 | K12879 | THO complex subunit 2 |
| STO1 | NCBP1 | K12882 | nuclear cap-binding protein subunit 1 |
| CBC2 | NCBP2 | K12883 | nuclear cap-binding protein subunit 2 |
| FAL1 | EIF4A3 | K13025 | ATP-dependent RNA helicase |
| SEC13 | SEC13 | K14004 | protein transport protein SEC13 |
| MLP1 | TPR | K09291 | nucleoprotein TPR |
| TIF5 | EIF5 | K03262 | translation initiation factor 5 |
| RPG1 | EIF3A | K03254 | translation initiation factor 3 subunit A |
| TIF34 | EIF3I | K03246 | translation initiation factor 3 subunit I |
| GCD2 | EIF2B4 | K03680 | translation initiation factor eIF-2B subunit delta |
| SUI3 | EIF2S2 | K03238 | translation initiation factor 2 subunit 2 |
| TIF1/TIF2 | EIF4A2 | K03257 | translation initiation factor 4A |
| TRZ1 | ELAC2 | K00784 | ribonuclease Z |
| NAM7 | UPF1 | K14326 | regulator of nonsense transcripts 1 |
| NMD2 | UPF2 | K14327 | regulator of nonsense transcripts 2 |

To measure the AFS of an alignment, we extract the GO terms from the Gene Ontology database[72]. We only consider aligned pairs in which both proteins have a GO term assigned. As shown in Figure 4.3, ModuleAlign yields better alignments than other methods in terms of AFS in categories BP, MF. There are similar results for category CC that is not presented here. ModuleAlign finds more functionally consistent protein pairs for all pairs of networks which facilitates the identification of conserved functional modules (see Figure 4.5).

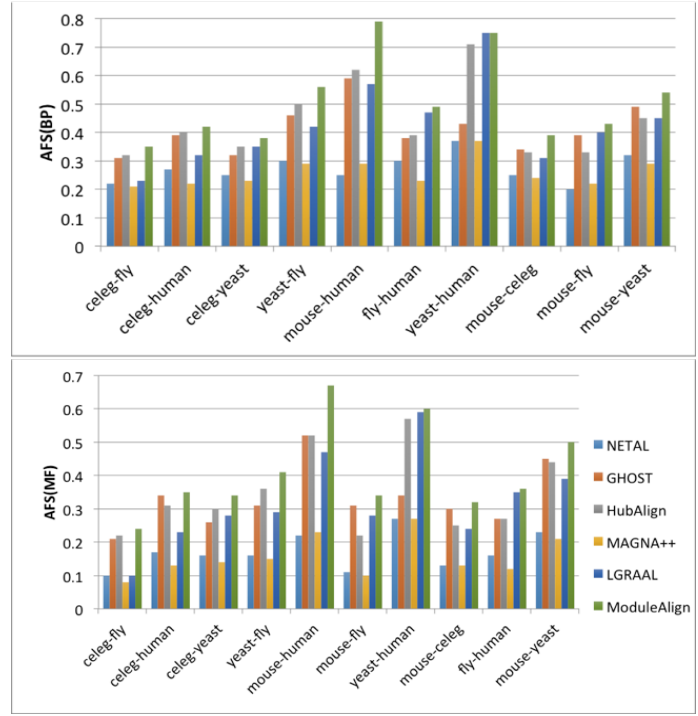

**Figure 4.3** Performance of all methods in terms of AFS in categories BP and MF.

Figure 4.4 shows that ModuleAlign produces alignments with larger *EC* and *LCCS* values than other methods except for NETAL. A larger *EC* value indicates that ModuleAlign can map densely connected proteins that potentially belong to similar structural or functional modules. While NETAL yields more aligned interactions than ModuleAlign, it is among the methods with lowest *AFS* indicating that many of the aligned proteins are not functionally similar (see Figure 4.3). We get similar results for $S^3$ score that are not shown here. For most of the instances ModuleAlign achieves the second-best score after NETAL, demonstrating ModuleAlign's ability to preserve sparse regions. Again, NETAL achieves a higher $S^3$ score, but at the cost of a very low *AFS* (see Figure 4.3). Overall, ModuleAlign is among the best aligners with respect to topological measures. More importantly, ModuleAlign aligns proteins with a substantially higher functional consistency than all competing methods and thus facilitates the transfer of functional annotations.

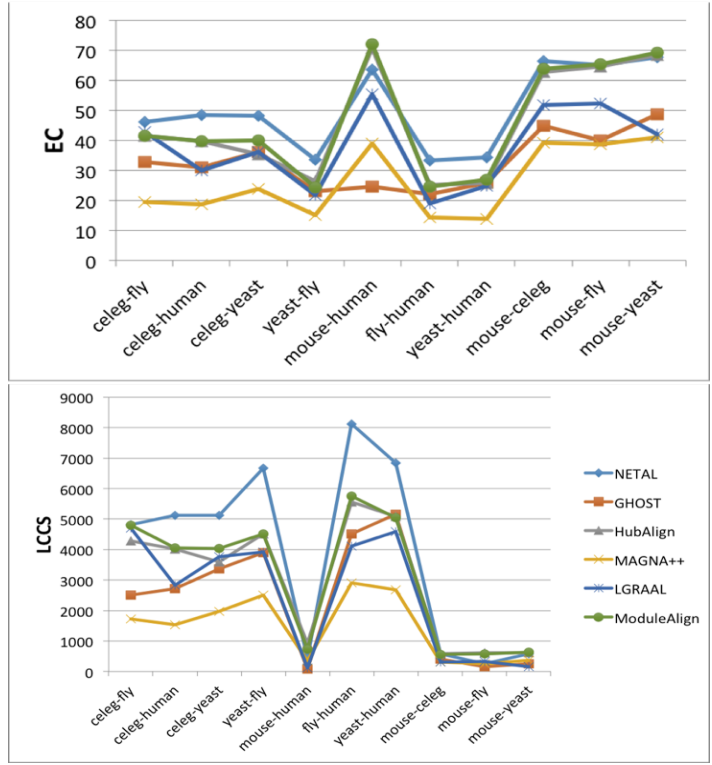

**Figure 4.4** Performance of ModuleAlign and competing methods in terms of EC and LCCS
Finding conserved sub-networks

### 4.3.2. Finding conserved sub-networks

A major application of network alignments is the identification of conserved sub-networks across two species. Figure 4.5 shows two conserved sub-networks between yeast and human detected by ModuleAlign that do not appear in the alignments produced by other methods. These sub-networks correspond to replication factor C complex and DNA replication, respectively (with $p-value < 10^{-7}$ in both species). In both sub-networks the proteins aligned by ModuleAlign have the same KO annotations, reflecting the high functional coherence between them. For yeast, some edges (shown by dotted lines) that are missing in our network are present in the String database[73], with experimental evidence at the highest confidence. This could suggest that ModuleAlign is tolerant to (partly) missing interactions, an extremely useful property in the context of incomplete PPI networks.

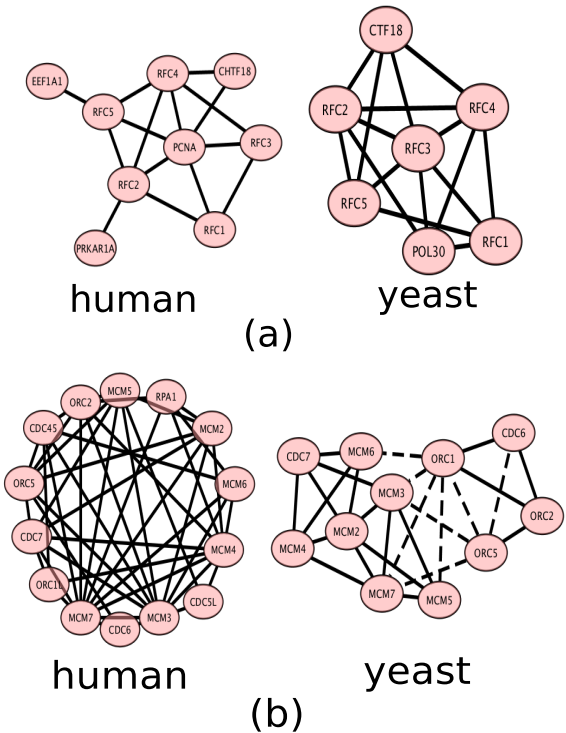

**Figure 4.5** The complexes detected by ModuleAlign in yeast-human alignment. The sub-networks are enriched for (a) replication factor C and (b) DNA replication.

### 4.3.3. Alignment of bacterial PPI networks

We also ran ModuleAlign to align the PPI networks of two bacterial species Escherichia coli (E. coli) and Campylobacter jejuni (C. jejuni) for which the most complete PPI networks among all bacteria exist. The E. coli and C. jejuni PPI networks have 1941 nodes, 3989 edges and 1111 nodes and 2988 edges, respectively[60, 74]. Escherichia coli is a model organism for studying fundamental cellular processes such as gene expression and signaling. Table 3 shows that ModuleAlign achieves a significantly higher precision and recall than all other methods. ModuleAlign also outperforms other methods in terms of AFS whose absolute values are small due to insufficient GO annotations of bacterial proteins. This again indicates that ModuleAlign can align proteins with consistent functions.

Table 4.4 $EC$, $S^3$ and $AFS$ of the alignments on the bacterial PPI networks.

|  | Precision | Recall | AFS (BP) | AFS (MF) | EC | $S^3$ |
|---|---|---|---|---|---|---|
| NETAL | 0 | 0 | 0.15 | 0.09 | 32.36 | 19.54 |
| GHOST | 0.05 | 0.04 | 0.19 | 0.14 | 22.79 | 15.14 |
| HubAlign | 0.31 | 0.24 | 0.25 | 0.22 | 24.65 | 16.51 |
| MAGNA++ | 0 | 0 | 0.24 | 0.18 | 24.83 | 19.32 |
| L-GRAAL | 0.11 | 0.09 | 0.12 | 0.10 | 24.61 | 17.83 |
| ModuleAlign | 0.37 | 0.30 | 0.31 | 0.28 | 25.95 | 16.92 |

### 4.3.4. Running time

ModuleAlign is among the fastest global alignment methods. We report the running times of all methods for the alignment of the human with the yeast and fly networks, which are the largest networks in our benchmark. On a 1400 MHz Linux system with 2GB RAM, it takes NETAL, HubAlign, MAGNA++, L-GRAAL, GHOST, and ModuleAlign 4, 15, 621, 64, 41, and 15 minutes, respectively, to align the yeast and human networks, and 7, 25, 757, 79, 50, and 30 minutes, respectively, to align the fly and human networks.

# Chapter 5

# Multiple PPI Network Alignment via Convex Optimization

## 5.1. Introduction

HubAlign and ModuleAlign are both designed for pairwise network alignment. But with the availability of more PPI networks, it becomes inevitable to align multiple networks. Although a few methods have been developed for multiple PPI network alignment, the alignment quality is still far from perfect and thus, new multiple network alignment methods are needed. Most of these current methods do not optimize alignment of all proteins simultaneously. Instead, they start from the best alignment between a subset of proteins and then gradually extend it by adding more proteins using a greedy strategy. This may impact alignment quality since errors introduced at an earlier stage cannot be fixed later.

In this chapter, we present a novel method, denoted as ConvexAlign [75], for joint alignment of multiple PPI networks by convex optimization of a scoring function composed of sequence similarity, topological score and interaction conservation score. It is NP-hard to optimize such a scoring function. We formulate this GNA problem as an integer program and relax it to a convex optimization problem, which enables us to simultaneously align all the PPI networks, without resorting to the widely-used seed-and-extension or progressive alignment methods. Then we use an alternating direction method of multipliers (ADMM) method to solve the relaxed convex optimization problem and optimize all the protein mappings together. In contrast to existing methods that generate multiple alignments in a greedy or progressive manner, our convex method optimizes alignments globally and enforces consistency among all pairwise alignments, resulting in much better alignment quality. Tested on both synthetic and real data, our experimental results show that ConvexAlign outperforms several popular methods in producing functionally coherent alignments. ConvexAlign even has a larger advantage over the others in aligning real PPI networks. ConvexAlign also finds a few conserved complexes among 5 species which cannot be detected by the other methods.

## 5.2. Methods

### 5.2.1. Scoring function for network alignment

Our goal is to find an alignment that maximizes the number of preserved edges and the number of matched orthologous (or functionally conserved) proteins. For this purpose, we use a node score for scoring matched proteins and an edge score for scoring matched interactions, respectively. For a pair of proteins, their node score is the combination of their topology score and sequence similarity score. We use a minimum-degree heuristic algorithm to calculate the topological score, which was used by us to develop a pairwise GNA method HubAlign. We use the normalized blast bit-scores for sequence similarity. Thus, node score $\text{node}(v_i, v_j)$, $v_i \in V_i$ and $v_j \in V_j$, is calculated as follows:

$$\text{node}(v_i, v_j) = (1 - \lambda_1) B(v_i, v_j) + \lambda_1 T(v_i, v_j),$$

where $\lambda_1$ controls the importance of the topology score relative to the blast score. The node score of multiple alignment $\mathcal{A}$, i.e. $f_{\text{node}}(\mathcal{A})$, sums the scores among all pairs of matched proteins:

$$f_{\text{node}}(\mathcal{A}) = \sum_{1 \leq i < j \leq N} \sum_{\mathcal{A}_k \in \mathcal{A}; v_i, v_j \in \mathcal{A}_k} \text{node}(v_i, v_j).$$

The edge score $f_{\text{interaction}}(\mathcal{A})$ measures interaction-preserving in an alignment $\mathcal{A}$. This score counts the number of interaction aligned between all pairs of networks:

$$f_{\text{interaction}}(\mathcal{A}) = \sum_{1 \leq i < j \leq N} \sum_{\mathcal{A}_k, \mathcal{A}_l \in \mathcal{A}; v_i, v_j \in \mathcal{A}_k; v'_i, v'_j \in \mathcal{A}_l} \delta((v_i, v'_i) \in E_i) \, \delta((v_j, v'_j) \in E_j),$$

where $\delta((v_i, v_i') \in E_i)$ is an indicator function. We aim to find the multiple alignment $\mathcal{A}$ that maximizes a combination of node and interaction scores where:

$$f(\mathcal{A}) = (1 - \lambda_2) f_{node}(\mathcal{A}) + \lambda_2 f_{interaction}(\mathcal{A}), \qquad (1)$$

where $\lambda_2$ describes the tradeoff. See Appendix for determination of λ1 and λ2 by cross-validation.

### 5.2.2. Integer and convex programming formulation

A one-to-one multiple network alignment is valid or feasible if the following condition (also called consistency property) is satisfied: for any three vertices $v_i$, $v_j$, and $v_k$ of three different networks, if $v_i$ is aligned to $v_j$ and $v_j$ aligned to $v_k$, then $v_i$ is aligned to $v_k$.

We encode a valid multiple alignment $\mathcal{A}$ using a binary matrix $Y = (Y_1; Y_2; \cdots; Y_N) \in \{0,1\}^{M \times K}$, where each block $Y_i$ describes the association between $V_i$ and $\mathcal{A}$ and $M = \sum_{i=1}^{N} |V_i|$. Each row of Y corresponds to one vertex and each column corresponds to one cluster of mutually-aligned vertices. In other words, $Y_i(v_i, \mathcal{A}_j) = 1$ if and only if $v_i \in V_i$ belongs to cluster $\mathcal{A}_j$. Let $\vec{1}$ be a vector of appropriate size with all elements being 1. Since Y is a one-to-one alignment, it shall satisfy the following constraints:

- Each row of Y has exactly one non-zero entry, i.e., $Y\vec{1} = \vec{1}$.
- Each column of Y has at most N non-zero entries, i.e. $Y\vec{1} \leq N\vec{1}$.
- Each column of $Y_i$ has at most one non-zero entry, i.e., $Y_i^T \vec{1} \leq \vec{1}$.

On the other direction, any binary matrix Y satisfying the above properties encodes a one-to-one alignment.

Although matrix Y provides a concise way to parametrize the space of valid multiple alignments, the objective function with Y is nonlinear and thus hard to optimize. Inspired by [76], we address this issue by introducing a correspondence matrix X where:

$$X = \begin{pmatrix} I_{|V_1|} & X_{12} & \cdots & X_{1N} \\ X_{12}^T & I_{|V_2|} & \cdots & X_{2N} \\ \vdots & \cdots & \ddots & \vdots \\ X_{1N}^T & \cdots & \cdots & I_{|V_N|} \end{pmatrix} = \begin{pmatrix} Y_1 \\ Y_2 \\ \vdots \\ Y_N \end{pmatrix} \cdot (Y_1^T \quad Y_2^T \quad \cdots \quad Y_N^T),$$

where each block $X_{ij} = Y_i Y_j^T$ is a binary matrix encoding the mapping between $V_i$ and $V_j$. That is, $X_{ij}(v_i, v_j) = 1$ if and only if $v_i$ and $v_j$ are aligned (i.e. in the same alignment cluster). Regarding the above equation, $f_{node}$ can be rewritten as follows:

$$f_{node}(\mathcal{A}) = \sum_{1 \leq i < j \leq N} \sum_{v \in V_i, v' \in V_j} node(v, v') X_{ij}(v, v') = \sum_{1 \leq i < j \leq N} \langle C_{ij} X_{ij} \rangle, \quad (2)$$

where matrix $C_{ij}$ encodes the values of $node(v, v')$ in its element.

To formulate $f_{interaction}$, we introduce indicator variables $y_{ij}(v_i, v_j, v_i', v_j')$ for edge correspondence:

$$y_{ij}(v_i, v_j, v_i', v_j') = X_{ij}(v_i, v_j) X_{ij}(v_i', v_j'), \quad \forall (v_i, v_i') \in E_i, (v_j, v_j') \in E_j, 1 \leq i < j \leq N. \quad (3)$$

$$f_{interaction}(\mathcal{A}) = \sum_{1 \leq i < j \leq N} \sum_{(v_i, v_i') \in E_i, (v_j, v_j') \in E_j} y_{ij}(v_i, v_j, v_i', v_j') = \sum_{1 \leq i < j \leq N} \sum_{(v_i, v_i') \in E_i, (v_j, v_j') \in E_j} \langle \vec{1}, y_{ij} \rangle, \quad (4)$$

where we use $y_{ij}$ to stack the indicators between $V_i$ and $V_j$.

The nonlinear constraint (3) can be replaced by the following inequalities [77]:

$$\forall v_j' \in V_j, \sum_{v_i' \in (v_i, v_i') \in E_i} y(v_i, v_j, v_i', v_j') \leq X_{ij}(v_i, v_j)$$

$$\forall v_i' \in V_i, \sum_{v_j' \in (v_j, v_j') \in E_j} y(v_i, v_j, v_i', v_j') \leq X_{ij}(v_i, v_j)$$

$$\forall v_j \in V_j, \sum_{v_i \in (v_i, v_i') \in E_i} y(v_i, v_j, v_i', v_j') \leq X_{ij}(v_i', v_j')$$

$$\forall v_i \in V_i, \sum_{v_j \in (v_j, v_j') \in E_j} y(v_i, v_j, v_i', v_j') \leq X_{ij}(v_i', v_j'). \quad (5)$$

It is easy to prove that (3) implies (5). On the other direction, considering that the coefficient of y is positive and we want to maximize (4), we can prove that (5) implies (3). We replace (3) by (10) to obtain linear constraints and summarize (10) in the matrix form:

$$B_{ij}y_{ij} \leq \mathcal{F}_{ij}(X_{ij}), \qquad (6)$$

where $B_{ij}$ is coefficient and $\mathcal{F}_{ij}$ is a linear operator that picks the corresponding element of $X_{ij}$ for each constraint. That is $\mathcal{F}_{ij}\left(X_{ij}(v_i, v_j)\right) \leq \langle P_{ij}, X_{ij}\rangle$ where $P_{ij}$ is a binary matrix with the same dimension as $X_{ij}$ and only one element $P_{ij}(v_i, v_j)$ is equal to 1.

Finally, by integrating (2), (4) and (6) we have the following convex program:

$$\text{maximize} \quad \sum_{1 \leq i < j \leq N} (1 - \lambda_2)\langle C_{ij}X_{ij}\rangle + \lambda_2\langle \vec{1}, y_{ij}\rangle$$

$$\begin{aligned}
\text{subject to} \quad & y_{ij} \in \{0,1\}^{|E_i| \times |E_j|}, \; B_{ij}y_{ij} \leq \mathcal{F}_{ij}(X_{ij}), & 1 \leq i < j \leq N \\
& X_{ij}\vec{1} \leq \vec{1}, \; X_{ij}^T\vec{1} \leq \vec{1}, \; X_{ij} \in \{0,1\}^{|V_i| \times |V_j|}, & 1 \leq i < j \leq N \\
& X \succcurlyeq 0, \; X_{ii} = I_{|V_i|}, & 1 \leq i \leq N \qquad (7)
\end{aligned}$$

The key constraint in above equation is the positive semi definite constraint $X \succcurlyeq 0$, which enforces the consistency among all pairwise alignments. Dropping it is equivalent to performing pairwise alignments in isolation.

### 5.2.3. Optimization via convex relaxation

It is NP-hard to directly optimizing (12) since the variables are integers. We may relax them to obtain a convex optimization problem that can be solved to globally optimal within polynomial time. The convex relaxation is tight, and a simple greedy rounding scheme can convert its factional solution to integral.

**Convex relaxation.** By relaxing $y_{ij}$ and $X_{ij}$ to real values between 0 and 1, we have the following convex program:

$$\begin{aligned}
\text{maximize} \quad & \sum_{1 \leq i < j \leq N} (1 - \lambda_2)\langle C_{ij} X_{ij}\rangle + \lambda_2 \langle \vec{1}, y_{ij}\rangle \\
\text{subject to} \quad & y_{ij} \geq 0, \; B_{ij} y_{ij} \leq \mathcal{F}_{ij}(X_{ij}), & 1 \leq i < j \leq N \\
& X_{ij}\vec{1} \leq \vec{1}, \; X_{ij}^T \vec{1} \leq \vec{1}, \; X_{ij} > 0, & 1 \leq i < j \leq N \\
& X \succcurlyeq 0, \; X_{ii} = I_{|V_i|}, & 1 \leq i \leq N \quad (8)
\end{aligned}$$

**Optimization strategy.** We employ ADMM (alternating direction of multiplier method) to solve the above convex program. The basic idea is to augment its Lagrangian and iteratively optimize a subset of variables while keeping the others fixed. This allows us to exploit structure patterns in the constraint set for effective optimization. For details of the derivations see [75].

**Rounding into an integer solution.** The factional solution to the above convex relaxation is usually tight, and a simple rounding strategy works reasonably well. Specifically, we collect all the protein pairs with an indicator value $X(u, v) > 0.05$ from large to small and place them into a sorted list $\mathcal{X}$. Starting from an alignment graph with an empty edge set, we scan through $\mathcal{X}$ in a decreasing order. For each scanned protein pair $(u, v)$ in $\mathcal{X}$, in the alignment graph we add an edge to connect this pair as long as such an addition does not violate the constraint that no protein in one network is aligned to two proteins in another network. At the very end when all pairs are scanned, we output all disconnected components of the alignment graph as the final output. Each component is a cluster of mutually-aligned proteins. Most of the resultant components are cliques. For the very few non-clique components, we just add some edges to make them cliques.

**Parameter selection.** For all the experiments in this paper, we set the parameters $\lambda_1 = 0.3$ and $\lambda_2 = 0.02$. These parameters are chosen via 10-fold cross-validation in optimizing the GO-term scores of the alignment between the mouse and worm networks. The weight factor for aligned interactions is small because: 1) there are many more aligned interactions than aligned nodes, so a small $\lambda_2$ may place the node and interaction scores at the similar scale; and 2) the topological score used in our scoring function already encodes some interaction information and thus, may overlap with the interaction score. We may increase $\lambda_2$ to favor other performance metrics such as the number of aligned interactions and the number of annotated clusters.

## 5.3. Results

We compare our algorithm ConvexAlign with several popular and publicly available methods IsoRankN, SMETANA, NetCoffee and BEAMS. We ran SMETANA and NetCoffee with their default parameters. For both BEAMS and IsoRankN, we set three different values for their parameter $\alpha = \{0.3, 0.5, 0.7\}$. We left other parameters of BEAMS at their default.

### 5.3.1. Test data

We use the PPI networks of H.sapiens (human), S.cerevisiae (yeast), Drosophila melanogaster (fly), Caenorhabditis elegans (worm) and Mus musculus (mouse) taken from IntAct [69]. The human network has 9003 proteins and 34935 interactions, the yeast network has 5674 proteins and 49830 interactions, the fly networks has 8374 nodes and 25611 interactions, the mouse network has 2897 proteins and 4372 interactions and the worm network has 4305 proteins and 7747 interactions. Only experimentally-validated PPIs are used.

We also use the NAPAbench [78] synthetic PPI networks. NAPAbench is a benchmark that contains PPI network families generated by different network models. We use the 8-way alignment dataset of this benchmark, which contains three network families each with 8 networks of 1000 nodes generated by one of the three network models. The 8-way alignment dataset simulates a network family of closely-related species, so this benchmark has very different properties as the above 5 real PPI networks.

### 5.3.2. Alignment quality on real data

**Topological quality.** Table 5.1 lists the topological evaluation of the alignments produced by different methods. The first four multi-rows show the results for the clusters consisting of proteins belonging to c = 2,3,4,5 species, respectively. In each multi-row, the top and bottom rows show c-coverage and the number of proteins in the clusters, respectively.

ConvexAlign has a larger c-coverage when c = 4,5 than the other methods except SMETANA and NetCoffee. However, as we show later, many of clusters generated by these two methods are not functionally conserved. The total coverage of BEAMS and IsoRank is better than the others because they produce many clusters composed of proteins from 2 or 3 species. These clusters cannot explain the data as well as clusters containing proteins from 4 or 5 species can. ConvexAlign has a better CI than all other methods except SMETANA. These conserved interactions may be very helpful in identifying the functional modules conserved among networks of different species. It is worth mentioning that most of the conserved interaction resulting from SMETANA may be spurious [1].

Table 5.1 Topological evaluation of output clusters by different alignment methods. IsoRankN and BEAMS are tested using three different values of their parameter $\alpha$.

|  | IsoRankN | | | SMETANA | NetCoffee | BEAMS | | | ConvexAlign |
|---|---|---|---|---|---|---|---|---|---|
|  | 0.3 | 0.5 | 0.7 | | | 0.3 | 0.5 | 0.7 | |
| c = 2 | 4625 | 4187 | 4670 | 1127 | 1424 | 5703 | 5274 | 5271 | 2856 |
|  | 11035 | 8356 | 11165 | 2718 | 2848 | 11406 | 11469 | 11465 | 5712 |
| c = 3 | 2259 | 2270 | 2304 | 1653 | 1739 | 2192 | 2557 | 2556 | 1833 |
|  | 8521 | 6810 | 8750 | 5808 | 5217 | 6576 | 8128 | 8118 | 5499 |
| c = 4 | 1023 | 731 | 944 | 2028 | 1980 | 1163 | 1141 | 1143 | 1190 |
|  | 5276 | 2924 | 4823 | 9531 | 7920 | 4652 | 4686 | 4701 | 4760 |
| c = 5 | 224 | 112 | 184 | 1622 | 1217 | 683 | 600 | 600 | 765 |
|  | 1417 | 560 | 1182 | 10342 | 6075 | 3915 | 3046 | 3044 | 3825 |
| c ≥ 2 | 8131 | 7291 | 8102 | 6430 | 6360 | 9741 | 9572 | 9570 | 6644 |
|  | 26249 | 18650 | 25920 | 28399 | 22070 | 26549 | 27329 | 27328 | 19796 |
| CI | 0.03 | 0.02 | 0.03 | 0.10 | 0.01 | 0.03 | 0.03 | 0.03 | 0.04 |

**Biological quality.** Table 5.2 shows the *AFS* separately for clusters composed of proteins in 3, 4, and 5 species in both categories BP and MF. The *AFS* obtained by ConvexAlign is 6 − 20% larger than the other methods. These results indicate that on average the clusters generated by ConvexAlign are functionally more consistent. That is, ConvexAlign outperforms the other methods in terms of not only the number of consistent clusters, but also the average GO semantic similarity. These results further confirm that ConvexAlign yields clusters with higher functional similarity in both categories MF and BP.

**Table 5.2** *AFS* comparison between ConvexAlign and the other methods

|  |  | IsoRankN | | | SMETANA | NetCoffee | BEAMS | | | ConvexAlign |
|---|---|---|---|---|---|---|---|---|---|---|
|  |  | 0.3 | 0.5 | 0.7 |  |  | 0.3 | 0.5 | 0.7 |  |
| $\overline{AFS}_{BP}$ | $c = 3$ | 0.83 | 1.02 | 0.86 | 0.74 | 1.03 | 1.60 | 1.63 | 1.63 | 1.74 |
|  | $c = 4$ | 0.69 | 0.97 | 0.72 | 0.68 | 0.99 | 1.63 | 1.61 | 1.60 | 1.79 |
|  | $c = 5$ | 0.75 | 1.01 | 0.72 | 0.85 | 1.16 | 1.66 | 1.67 | 1.67 | 1.71 |
| $\overline{AFS}_{MF}$ | $c = 3$ | 0.80 | 0.94 | 0.80 | 0.69 | 0.99 | 1.40 | 1.33 | 1.34 | 1.54 |
|  | $c = 4$ | 0.83 | 1.02 | 0.86 | 0.74 | 1.03 | 1.60 | 1.63 | 1.63 | 1.74 |
|  | $c = 5$ | 0.86 | 1.06 | 0.86 | 0.94 | 1.18 | 1.68 | 1.68 | 1.68 | 1.74 |

Table 5.3 provides the functional consistency measures of the alignments generated by competing method. The first four multi-rows show the quality of the clusters composed of proteins from c = 2,3,4,5 species. In these multi-rows, the top and middle rows show the number of consistent and annotated clusters, respectively, and the bottom row shows specificity. Regardless of c, ConvexAlign[75] outperforms the other methods in terms of specificity and the number of consistent clusters. At the same time, ConvexAlign generates fewer annotated clusters that BEAMS when c = 2,3,4. Although SMETANA and NetCoffee generate a larger number of clusters for c = 4,5 than ConvexAlign, their clusters are not very functionally consistent. The fifth row shows ConvexAlign has much higher specificity than the others when all the resulting clusters c ≥ 2 are considered. These results suggest that ConvexAlign finds more functionally consistent clusters, not only by generating small clusters (i.e. c = 2,3) but more importantly large clusters (i.e. c = 4,5). These clusters (especially when c = 4,5) are very valuable because they may provide useful information about the orthology relationship among the proteins of all species. Moreover, these clusters can be very useful for identifying conserved sub-networks as well as predicting the function of unannotated proteins. ConvexAlign yields a COI/CI ratio around 60% that is 1.44 times larger than the second-best ratio by BEAMS. This result may indicate that ConvexAlign can identify conserved interactions between orthologous proteins. It also suggests that although SMETANA has the largest CI, many of those conserved interactions are possibly false and formed by non-orthologous proteins. ConvexAlign also outperforms other methods in terms of MNE and sensitivity.

**Table 5.3** Functional consistency of output clusters. Note that for $MNE$, the smaller the better; while for the other measure, the larger the better.

|  |  | IsoRankN | | | SMETANA | Net Coffee | BEAMS | | | Convex Align |
|---|---|---|---|---|---|---|---|---|---|---|
|  |  | 0.3 | 0.5 | 0.7 |  |  | 0.3 | 0.5 | 0.7 |  |
| $c = 2$ | consistent | 906 | 1259 | 919 | 295 | 495 | 1539 | 1568 | 1569 | 1914 |
|  | annotated | 3614 | 2862 | 3646 | 931 | 26 | 3486 | 3456 | 3452 | 2326 |
|  | specificity | 0.25 | 0.44 | 0.25 | 0.39 | 0.53 | 0.44 | 0.45 | 0.45 | 0.82 |
| $c = 3$ | consistent | 203 | 466 | 231 | 188 | 462 | 1003 | 1084 | 1084 | 1155 |
|  | annotated | 2160 | 2153 | 2210 | 1556 | 1640 | 2119 | 2442 | 2441 | 1741 |
|  | specificity | 0.09 | 0.22 | 0.10 | 0.12 | 0.28 | 0.47 | 0.44 | 0.44 | 0.66 |
| $c = 4$ | consistent | 41 | 106 | 54 | 170 | 406 | 606 | 624 | 624 | 661 |
|  | annotated | 1020 | 723 | 942 | 2019 | 1640 | 1159 | 1136 | 1138 | 1079 |
|  | specificity | 0.04 | 0.15 | 0.06 | 0.08 | 0.25 | 0.52 | 0.55 | 0.55 | 0.61 |
| $c = 5$ | consistent | 14 | 19 | 9 | 183 | 406 | 383 | 359 | 359 | 493 |
|  | annotated | 224 | 112 | 184 | 1621 | 1955 | 683 | 600 | 600 | 763 |
|  | specificity | 0.06 | 0.17 | 0.05 | 0.11 | 0.21 | 0.56 | 0.60 | 0.60 | 0.65 |
| $c \geq 2$ | Specificity | 0.17 | 0.32 | 0.17 | 0.14 | 0.29 | 0.48 | 0.48 | 0.48 | 0.71 |
|  | COI | 88 | 188 | 127 | 480 | 553 | 1237 | 1311 | 1305 | 1668 |
|  | COI/CI | 0.02 | 0.13 | 0.03 | 0.04 | 0.21 | 0.40 | 0.41 | 0.41 | 0.59 |
|  | MNE | 2.15 | 2.19 | 2.14 | 2.44 | 2.39 | 1.97 | 1.95 | 1.95 | 1.93 |
|  | Sensitivity | 0.45 | 0.46 | 0.45 | 0.36 | 0.22 | 0.33 | 0.31 | 0.37 | 0.51 |

### 5.3.3. Alignment quality on synthetic data

This section explains the results on the NAPAbench benchmark. Figure 5.1 shows the number of consistent clusters generated by different methods and their specificity on clusters composed of proteins in c = 2,3,4,5,6,7,8 species, respectively. In terms of the number of consistent clusters, ConvexAlign is slightly better than the second-best method BEAMS regardless of c, but much better than the others. In terms of specificity, ConvexAlign has a much larger advantage over the other methods when c = 4,5,6,7. When c = 8, ConveAlign is slightly better than BEAMS, but much better than the others. These results indicate that ConvexAlign aligns proteins in a functionally consistent way, without generating too many spurious clusters in which the proteins appear to be unrelated.

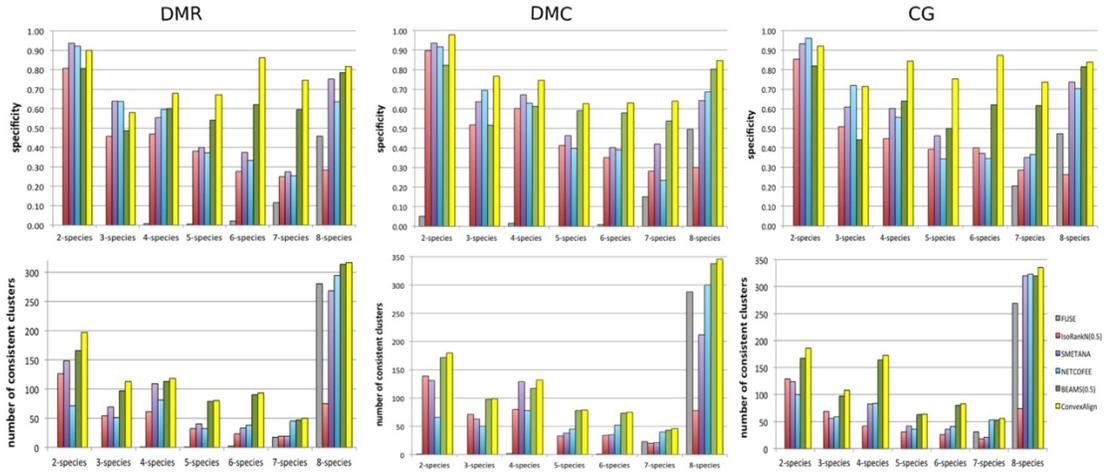

**Figure 5.1** Specificity and the number of consistent clusters generated by the competing methods for different c on synthetic data. Only the best performance for IsoRankN and BEAMS is shown.

Figure 5.2 shows that ConvexAlign outperforms all the other methods in terms of both MNE and COI.

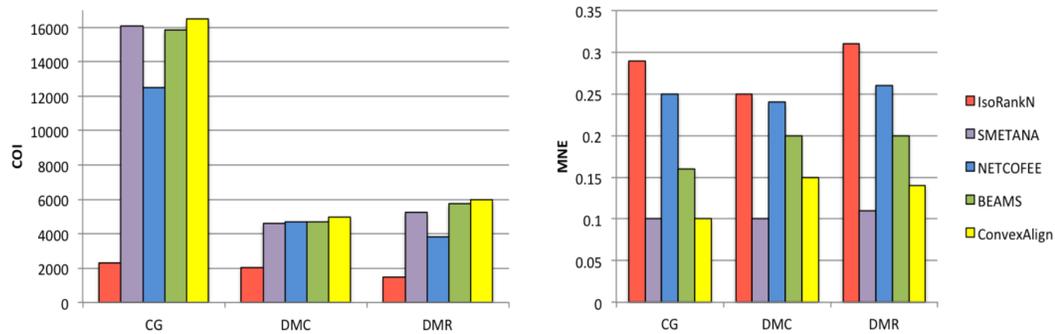

**Figure 5.2** COI and MNE of the clusters generated by the competing methods on synthetic data. Only the best performance for IsoRankN and BEAMS is shown.

### 5.3.4. Finding conserved sub-networks

One of the applications of network alignment is to reveal sub-networks conserved across the species. These sub-networks may be helpful for extracting biological information that cannot be inferred from sequence similarity alone. Figure 5.3 shows one conserved complex detected by ConvexAlign among the five species: human, yeast, fly, mouse and worm, but not appearing in the alignments generated by other methods. This complex is enriched for DNA replication. In this Figure, the interactions in IntAct are displayed in solid lines. For fly, mouse and worm, some edges (shown by dotted lines) are missing in IntAct but present in the STRING database [73] with experimental evidence at the highest confidence. Note that our input networks consist of interactions only from IntAct but not STRING. This suggests that ConvexAlign can predict missing interactions. We use PANTHER [79] to check if the aligned nodes are orthologous proteins. Most of the aligned proteins are shown to be least divergent orthologs.

### 5.3.5. Running time

ConvexAlign is computationally efficient compared to the other methods. Tested on the alignment of the networks of five species, it takes ConvexAlign, IsoRankN, BEMAS, FUSE, NetCoffee, and SMETANA 480, 1129, 900, 780, 15, and 37 minutes, respectively, to terminate. Although NetCoffee has the smallest running time, it does not yield alignments with significant functional consistency.

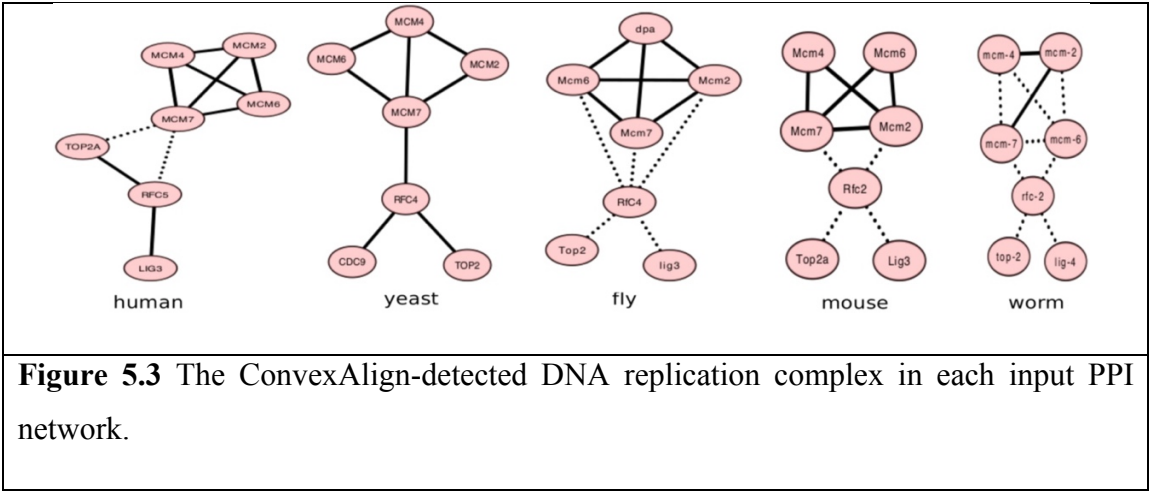

**Figure 5.3** The ConvexAlign-detected DNA replication complex in each input PPI network.

# Chapter 6

# Inference of Tissue-specific Networks

## 6.1. Introduction

It is well known that the majority of human genes have varying levels of expression in different tissues [80] that reflect variations in functional modules, and characteristics of biological pathways for each particular tissue. One way to understand the tissue-specific molecular mechanism behind these gene expressions is to generate co-expression networks, in which two genes are connected if their expression levels behave similarly. The correlation between a pair of expression profiles may imply that their corresponding genes are functionally related [81, 82]. Tissue-specific CENs provide useful information for prediction of tissue-specific functional modules, prioritization of disease causing genes and also identification of drug targets [83]. Such information cannot be revealed by a global co-expression network because it neglects the critical aspect of biology, that proteins play a different role among diverse tissues.

There are two major challenges in constructing tissue-specific networks. First, tissue-specific expression profiles usually have too few samples compared to the number of genes that makes it difficult to infer an accurate network. We may mitigate this problem by exploiting the similarities between different tissues. On the other hand, biological networks are known to be scale free, meaning that some nodes have many interactions while other nodes have only a few. One solution may be to penalize the edges adjacent to a hub less than the other edges thereby penalizing nodes instead of edges.

We use Gaussian graphical models with a new regularization term to model multiple functionally related tissues with a set of multivariate Gaussian distributions, where the inverse covariance matrix of each distribution encodes the CEN of each tissue. Our method is based upon the observation that biological networks are scale-free and functionally similar tissues are more likely to share similar interactions. We use a group lasso penalty to exploit the similarity between the related tissues, as well as a node penalty to differentiate between a hub and non-hub node that helps to find a scale-free network structure. By combination of these two penalties we provide a unique approach to mitigate the shortcoming that many tissues have few samples, allowing to infer a precise network for each of them.

## 6.2. Method

We represent the gene expression profile of a tissue t by matrix $X^t \in \mathcal{R}^{n \times m}$ -- each row represents an experiment and each column a gene. We assume each experiment $x_i^t$ ($1 \leq i \leq n$) is drawn from a normal distribution $\mathcal{N}(\mu^t, \Sigma^t)$, with mean vector $\mu^t = (\mu_1, \ldots, \mu_m)$ and covariance matrix $\Sigma^t = (\sigma_{ij})_{(1 \leq i,j \leq m)}$. Let $(\Sigma^t)^{-1}$ denote the precision matrix. The gene co-expression network for a tissue $t$ is represented by an undirected graph $G^t = (V^t, E^t)$ where $V^t$ is the set of vertices (genes) and $E^t$ the set of edges (correlations between genes). $|V^t|$ and $|E^t|$ are respectively the number of nodes and edges of the network.

### 6.2.1. Modeling tissue-specific network

We model the tissue-specific CENs by a multivariate Gaussian distribution. The log-likelihood of the data can be written as:

$$log\left(p(\{X^i\}|S)\right) = \sum_{t=1}^{K} n_t \left(log(det(S^t)) - tr(S^t \hat{\Sigma}^t)\right) + C,$$

where $C$ is a constant, $K$ is the number of tissues and $S = \{S^1, S^2, \ldots, S^K\}$ is the set of precision matrices to be estimated and $\hat{\Sigma} = \{\hat{\Sigma}^1, \hat{\Sigma}^2, \ldots, \hat{\Sigma}^K\}$ is the set of covariance matrices. The non-zero entries in the precision matrices correspond to the interactions among genes. Because we predict the CENs of a pair of similar tissues at a time, $K$ is set to 2. Let $p(S)$ denote a prior on the precision matrices, which can be used for different purposes (e.g. sparsity of the precision matrix). We estimate the precision matrices by Maximum a posteriori (MAP) as follows:

$$max_S log\left(p(\{X^i\}|S)\right) + log\, p(S).$$

Maximizing above equation with respect to S, yields the estimates $\hat{S} = \{\hat{S}^1, \hat{S}^2, ..., \hat{S}^K\}$ for the precision matrices. To define the regularization term, we reshape the matrix $S$ into sum of two matrices $V$ and $Z$ as follows:

$$S = Z + (V + V')/2,$$

where $Z$ is a sparse symmetric matrix and represents edges between non-hub nodes, and $V$ is a matrix whose columns are entirely zero or have many non-zero elements where a non-zero column represents a hub [84]. Notice that $V'$ is the transpose of $V$. Moreover, we apply a group lasso penalty to the $(i,j)$ element across all $K$ precision matrices to encourage a similar pattern of sparsity across all the precision matrices of related tissues. We propose the following regularization term to jointly estimate the scale-free networks of related tissues:

$$\log p(S) = -(\lambda_1 \sum_{t=1}^{K} \|Z^t\|_1 + \lambda_2 \sum_{t=1}^{K} \|V^t\|_1 + \lambda_3 \sum_{t=1}^{K} \sum_{j=1}^{m} \|V_j^t\|_2^2 +$$

$$\lambda_4 \sum_{i \neq j} \sqrt{\sum_{t=1}^{K} (S_{ij}^t)^2}),$$

$$subject\ to\ S = Z + \frac{V+V'}{2}, \quad (3)$$

where $\lambda_1, \lambda_2, \lambda_3$ and $\lambda_4$ are tuning parameters, $\lambda_1, \lambda_2$ and $\lambda_3$ control the sparsity of the precision matrix and $\lambda_4$ encourages the precision matrices to share certain characteristics, such as the pattern of nonzero elements. This means that if there is some similarity between the input tissues, only those false positive edges would be eliminated by JointNet, while those true positive tissue-specific interactions would be retained. For simplicity, we assume that $n_1 = n_2 = \cdots = n_K = n$.

### 6.2.2. Optimization strategy

We solve the optimization problem using an alternating directions method of multipliers (ADMM) algorithm [85]. Note that since ADMM utilizes eigenvalue decomposition, this partitioning significantly reduces the running time. We select tuning parameters for JointNet using an approximation of the Akaike Information Criterion (AIC) [86] as follows:

$$AIC(\lambda_1, \lambda_2, \lambda_3, \lambda_4) = \sum_{t=1}^{K} \left(- \log \det \hat{S}^t + tr(\hat{S}^t \hat{\Sigma}^t) + 2I_{\hat{S}^t}\right),$$

where $\hat{S}^t$ is an element of the set of estimated inverse covariance matrices based on tuning parameters $\lambda_1, \lambda_2, \lambda_3$ and $\lambda_4$, and $I_{\hat{S}^t}$ is the number of non-zero elements in $\hat{S}^t$. Then, a grid search can be applied to select the tuning parameters minimizing the $AIC(\lambda_1, \lambda_2, \lambda_3, \lambda_4)$ score.

## 6.3. Results

### 6.3.1. Test data

The tissue-specific networks were predicted for five pair tissues that are known to be functionally related [87]. Table 6.1 shows the selected tissues and their samples size. Since JointNet does not depend on how pairs of tissues are formed, it would also work with any measure of tissue similarity such as mean levels of gene expression. The gene expression profiles for the tissues are taken from a dataset of a global map of human gene expression measurements for 12204 genes [88].

Table 6.1 Distance values between each pair of tissues. The smaller values indicate that tissues are more similar

| Pair tissue | Distance | Sample size |
|---|---|---|
| Superior (Area 9)/primary (Area 4) | 0.0193 | 12/15 |
| Atrial myocardium/cardiac ventricle | 0.0058 | 18/13 |
| Skeletal muscle /quadriceps muscle | 0.0136 | 9/7 |
| Tonsil/lymph node | 0.0471 | 10/10 |
| Eye/Thyroid | 10.0451 | 6/7 |

### 6.3.2. Comparison with other methods

We compare JointNet with four other popular and publicly available algorithms ARACNE [89], CLR [90], MRNET [91] and GeCON [92] using functional similarity based on GO terms. For this purpose, following [80], we select a subset of highly expressed genes in each tissue, which are annotated by at least one biological process. Pairs of proteins that are co-annotated to similar biological processes are more likely to be connected to each other [80]. Here, the biological similarity between the proteins is calculated using the Schlicker's similarity, based on the Resnik ontological similarity [41]. We compute two sets of baselines using ARACNE, CLR, MRNET and GeCON, (1) learning a separate network for each tissue independently, or (2) learning a single network for each pair of similar tissues by first concatenating their corresponding gene expression profiles and treating them as independent observations from the same distribution. For each inferred network, we select a subset of top ranked interactions at three different thresholds (5000, 10000 and 15000) and compute the fraction of interactions in the subset for which the functional similarity is more than 0.5. We refer to this score as the functional similarity score, quantifying how precise the networks are at capturing the biological properties of different tissues. Table 6.2 and Table 6.3 respectively show the performance of the methods for the first and second baselines.

The results show that JointNet predicts tissue-specific CENs more accurately than either baseline in all tissues, which means it can identify the interactions between genes that participate in the similar biological processes. This suggests that the better performance of JointNet is based on better modeling of CENs and not simply due to increasing the amount of information. Moreover, interestingly, comparing Table 6.2 with Table 3 suggests that concatenating the expression profiles and inferring a single network for a pair of tissues, is an inferior approach to learning a separate network for each.

**Table 6.2** Functional similarity score of methods when the network of each tissue is inferred separately.

| Tissue | #top ranked edges | CLR | ARCHNE | MRNET | GeCON | JointNet |
|---|---|---|---|---|---|---|
| Eye | 5000 | 66.61 | 63.27 | 59.45 | 18.32 | 67.71 |
|  | 10000 | 65.99 | 65.39 | 64.43 | 45.87 | 66.63 |
|  | 15000 | 64.72 | 65.47 | 64.99 | 44.87 | 66.45 |
| Superior | 5000 | 62.98 | 64.95 | 61.30 | 66.92 | 69.37 |
|  | 10000 | 63.73 | 62.27 | 62.62 | 57.68 | 68.35 |
|  | 15000 | 64.33 | 61.89 | 63.96 | 60.97 | 67.32 |
| Cardiac | 5000 | 66.06 | 65.38 | 65.75 | 64.10 | 76.15 |
|  | 10000 | 66.85 | 64.69 | 66.14 | 58.13 | 77.22 |
|  | 15000 | 65.85 | 66.58 | 66.37 | 61.71 | 77.43 |
| Atrial | 5000 | 65.71 | 64.72 | 67.03 | 60.70 | 77.03 |
|  | 10000 | 65.91 | 64.58 | 66.19 | 56.84 | 77.90 |
|  | 15000 | 65.40 | 65.35 | 66.29 | 62.02 | 79.46 |
| Thyroid | 5000 | 57.25 | 60.53 | 60.30 | 17.25 | 66.73 |
|  | 10000 | 62.05 | 62.51 | 65.93 | 47.21 | 67.11 |
|  | 15000 | 63.72 | 63.16 | 65.60 | 48.58 | 67.56 |
| Tonsil | 5000 | 67.62 | 65.43 | 66.70 | 67.22 | 68.47 |
|  | 10000 | 67.31 | 65.99 | 65.95 | 55.57 | 67.39 |
|  | 15000 | 65.77 | 66.85 | 66.60 | 60.28 | 67.52 |
| Primary | 5000 | 64.73 | 62.21 | 64.28 | 67.07 | 67.97 |
|  | 10000 | 64.78 | 64.57 | 64.92 | 56.84 | 67.21 |
|  | 15000 | 54.05 | 55.12 | 53.98 | 61.46 | 66.20 |
| Lymphnode | 5000 | 66.43 | 62.73 | 63.57 | 60.73 | 67.47 |
|  | 10000 | 66.22 | 62.65 | 64.86 | 57.63 | 66.27 |
|  | 15000 | 66.53 | 66.28 | 66.42 | 52.18 | 68.69 |
| Skeletal | 5000 | 65.24 | 64.20 | 62.20 | 16.25 | 65.79 |
|  | 10000 | 61.66 | 62.97 | 62.45 | 45.82 | 64.14 |
|  | 15000 | 63.89 | 63.55 | 63.38 | 44.87 | 63.87 |
| Quadriceps | 5000 | 62.50 | 61.44 | 64.08 | 17.25 | 66.43 |
|  | 10000 | 61.29 | 61.99 | 63.37 | 45.87 | 65.46 |
|  | 15000 | 63.45 | 62.39 | 63.40 | 44.87 | 66.44 |

Table 6.3 Functional similarity score of the methods when we combine the gene expression profile of two tissues and predict a single network for them. We also list our results in this table for the readers. For JointNet, the average of functional similarity scores for each pair of tissue is shown.

| Tissue | #top ranked interactions | CLR | ARACNE | MRNET | GeCON | JointNet |
|---|---|---|---|---|---|---|
| Eye-Thyroid | 5000 | 62.41 | 65.33 | 64.56 | 17.25 | 67.22 |
| | 10000 | 62.19 | 63.30 | 62.75 | 45.87 | 66.87 |
| | 15000 | 63.91 | 65.05 | 64.10 | 44.87 | 67.00 |
| Superior-Primary | 5000 | 64.80 | 65.45 | 64.36 | 67.07 | 68.67 |
| | 10000 | 64.01 | 63.92 | 64.53 | 56.80 | 67.78 |
| | 15000 | 64.98 | 63.49 | 64.47 | 60.97 | 66.76 |
| Atrial-Cardiac | 5000 | 62.85 | 66.62 | 63.74 | 64.70 | 76.59 |
| | 10000 | 63.28 | 65.61 | 63.46 | 56.84 | 77.56 |
| | 15000 | 63.16 | 65.54 | 64.25 | 62.02 | 78.44 |
| Tonsil-Lymphnode | 5000 | 62.59 | 65.17 | 63.09 | 60.73 | 67.97 |
| | 10000 | 63.96 | 64.11 | 62.68 | 57.63 | 66.83 |
| | 15000 | 63.99 | 63.89 | 63.91 | 62.18 | 68.10 |
| Skeletal-Quadricep | 5000 | 61.65 | 60.00 | 62.23 | 17.25 | 66.11 |
| | 10000 | 63.21 | 63.86 | 61.24 | 45.87 | 64.80 |
| | 15000 | 63.93 | 62.96 | 63.47 | 44.87 | 65.15 |

### 6.3.3. Comparison with independently-predicted network

To compare our jointly-inferred tissue-specific networks with those inferred independently for each tissue (i.e. $\lambda_4 = 0$), we use the gene expression profile of each pair of similar tissues to reconstruct their tissue-specific networks using different values for $\lambda_4 \in$ Error! Bookmark not defined. (Other tuning parameters are fixed). We then compare the performance of the algorithm in terms of the biological functional similarity score. Figure 6.1 shows when two tissues are very similar (e.g. cardiac and atrial) a larger $\lambda_4$ is preferable to promote network similarity. On the other hand, when two tissues are not very similar (e.g. eye and thyroid), a smaller $\lambda_4$ is favored. Moreover, these results suggest that using the similarity between related tissues lead to more biologically meaningful tissue-specific co-expression networks.

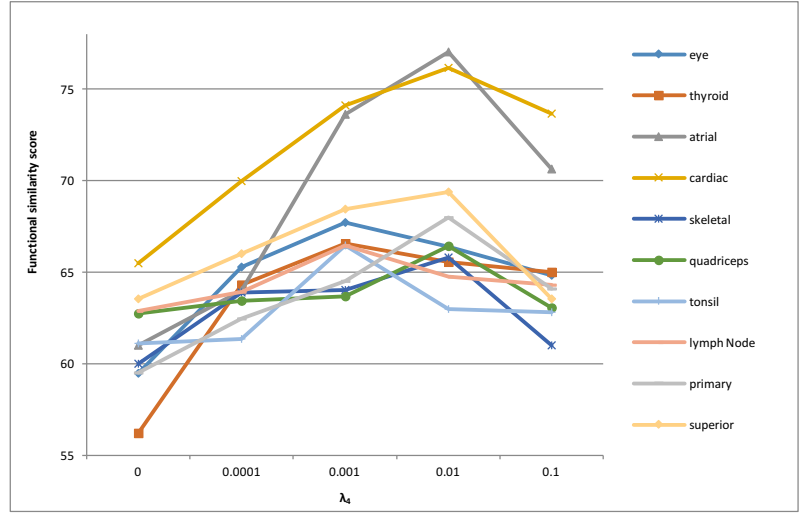

**Figure 6.1** Network quality measured by functional similarity score with respect to different $\lambda_4$ values.

### 6.3.4. System-level analysis of the inferred networks

**Degree distribution.** In the first experiment, the degree distribution of the nodes was plotted on a log-log scale. Biological networks are known to have a scale-free topology, which implies a power-law pattern for degree distribution [93]. Figure 6.2 shows the degree distribution plot of some of the inferred tissue-specific CENs. As expected, we observe a power-law pattern on degree distribution for all predicted tissue-specific CENs. We fit a power law function of the form $f(x) = ax^{-\gamma}$ to the degree distribution to see how well it fits a scale free distribution and report the coefficient of determination (i.e. $R^2$). Our result is in line with a recent survey of existing results on scaling of biological networks, which showed that biological networks have exponents that are between 1 and 2; that is, $1 \leq \gamma \leq 2$ [94].

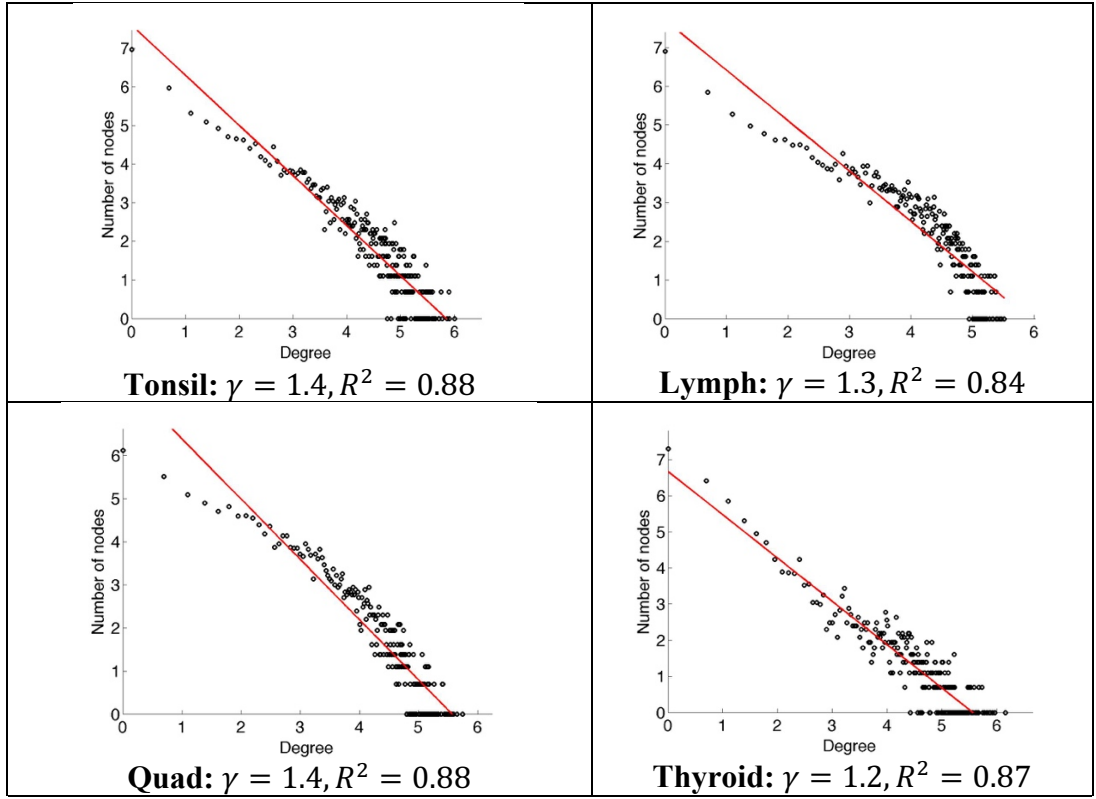

**Figure 6.2** Degree distribution on log scale. Red line is the fitted power law.

**Tissue specific transcription factors.** In the second experiment, to strengthen our analysis, we rely on a set of tissue-specific transcription factors (TF) from [95] that are known to act as module regulators responsible for tissue-specificity. Co-expression networks can involve many unique transcription factors whose activities are different across different tissues [95]. It is observed that average degree of the tissue specific TFs are usually higher than the average degree of the other nodes in their corresponding tissue, so we study the degree of TFs in the constructed CENs. For this purpose, we define ratio ρ as follows:

$$\rho = \frac{\bar{d}_{TF} - \bar{d}}{\bar{d}} \times 100,$$

where $\bar{d}_{TF}$ is the average degree of tissue specific transcription factors and $\bar{d}$ is the average degree of other nodes in a network. A larger ratio indicates that most of the transcription factors have a relatively high degree compared to the other nodes, while smaller values mean that TFs are not well distinguished from the rest of nodes in the network. Figure 6.3 reports the values of ρ in each tissue-specific CEN that were inferred jointly using JointNet or inferred independently. The results show that in eight out of ten tissues the ratio is much larger in networks resulting from JointNet compared to the networks inferred independently. This may indicate that in the CENs inferred by JointNet, most of the transcription factors have a higher degree compared to the other nodes and therefore, co-regulate with many other genes to exert their effects on the tissue. Moreover, in most of the networks inferred independently, the value of ρ is less than zero which means the average degree of tissue-specific TFs is much less than the average degree of the other nodes in the corresponding tissue. Once again, this confirms that the accuracy of the tissue-specific CENs is because of better modeling of commonalities and dissimilarities among tissues.

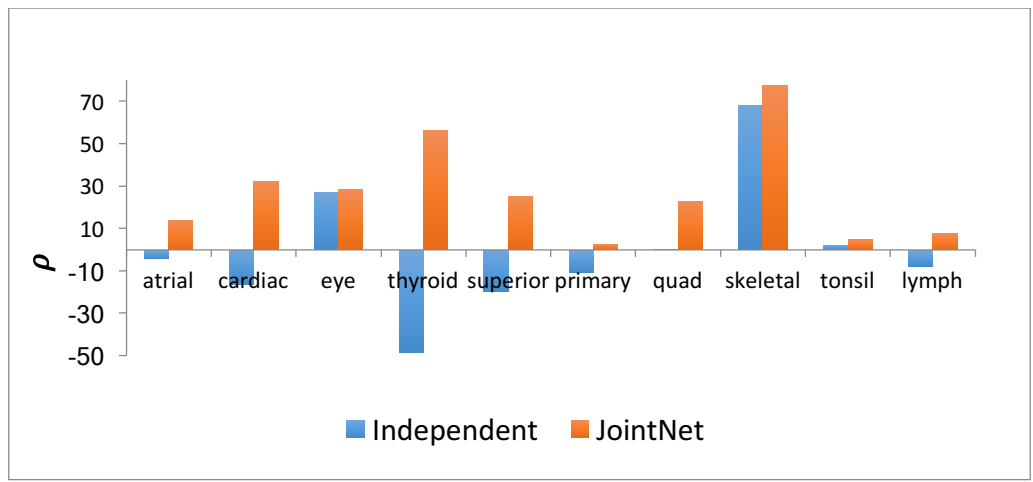

**Figure 6.3** Comparison of the tissue specific TFs based on ρ

# Chapter 7

# Predicting Protein-Protein Interactions through Sequence-based Deep Learning

## 7.1. Introduction

In this section, we present a novel deep learning framework, DPPI, to model and predict direct physical protein-protein interactions from primary amino acid sequences (Figure 7.1-A), Online Methods, details can be seen in Supplementary Notes). Our deep learning framework can (1) efficiently handle massive amounts of training data by using parallelized GPU-based learning algorithms, (2) flexibly train models for different applications without significant parameter tuning, and (3) robustly capture complex and non-linear relationships involved in protein interactions. For training, DPPI takes as input a pair of protein sequences and an experimentally determined binding score. Binding scores can be a binary label (0=non-binding, and 1=binding) or real values (e.g. enrichment value or dissociation constant, $K_d$).

We introduce three major aspects to our novel deep learning framework. First, DPPI leverages a large corpus of unsupervised data to build a robust profile representation of each protein sequence (Figure 7.1-B)[18, 96]. DPPI uses a sliding window to look at patches of linear peptide sequences and then builds probabilistic profiles using the known proteome to characterize the peptide sequence. Thus, instead of the raw sequence, a high dimensional position-specific profile representation of the sequences is given to the model. This enables non-identical but homologous proteins, such as from human and mouse, to have similar sequence profile representations. Furthermore, our use of a sliding window also enables DPPI to effectively operate on sequences of variable lengths by training on all combinations of patches. From a deep-learning perspective, while most patches between two proteins may not be involved in direct interactions, these "random" combinations are a form of data augmentation that improves model robustness by injecting and training with random noise[97, 98].

Second, DPPI exploits a Siamese-like convolutional neural network architecture. Siamese neural networks are useful for tasks that involve learning a complex relationship between two entities[99]. Our architecture contains two identical sub-networks, which share the same configuration, parameters, and weights. Within each sub-network, the convolutional module (Figure 7.1-C) convolves a peptide profile with specific parameters (filter). Here, the rectification layer clamps all the negative values to zero to introduce non-linearity, and the pooling stage computes the average of each filter's rectified response across the profile. Each sub-network produces a high-dimensional representation of a single protein sequence. Importantly, parameter sharing guarantees that two similar sequences cannot be transformed by their respective sub-networks to very different feature representations.

Third, we introduce a novel random projection module. The values of the last convolutional module are randomly projected into a subspace using a pair of (pseudo-orthogonal) random weight vectors (Figure 7.1-D). The pair of random untrained weight vectors are generated only once before training, and gives the model its unique ability to accommodate both homodimeric (identical sequences) and heterodimeric (dissimilar sequences) proteins interactions. The random weights are reversed between the two-networks, which induces symmetry and enables the model to ignore the order of the input profiles to either sub-network. Lastly, the final vectors are combined as a high-dimensional feature representation of the protein pair and are used to calculate an interaction score. For predicting and testing the interaction between two proteins the maximum interaction score over all patches is reported.

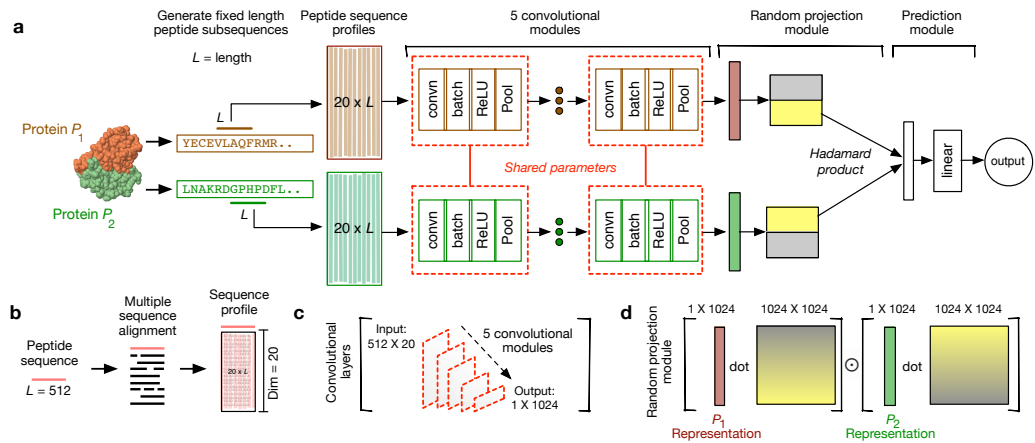

**Figure 7.1** DPPI architecture

## 7.2. DPPI design

DPPI has three main modules: a convolutional module, a random projection module and a prediction module. The core module of our model is the convolutional module, which learns a set of filters that is responsible for detecting patterns in protein sequence.

The convolutional module maps a pair of protein sequences to a representation that is useful for predicting protein-protein interactions. The role of random projection module is to take the representation learned by the convolutional module for each of the two proteins and use two different mappings to project these representations onto two different spaces. The prediction module takes the representations generated by the random projection module and outputs a score that is then used to predict existence of an interaction between two given proteins.

### 7.2.1. Input

DPPI takes a pair of protein profiles S and S′ and outputs a binary value f(S, S′) indicating whether the corresponding proteins interact. The profiles for the proteins were made by Blast using the entire uniprot, where redundancy was reduced to 80% maximum pairwise sequence similarity. For a given primary protein sequence with length n we generate a n×20 array S, called profile, as follows:

$$S = \begin{bmatrix} s_{1,1} & \cdots & s_{1,20} \\ \vdots & & \vdots \\ s_{i,1} & \vdots & s_{i,20} \\ \vdots & & \vdots \\ s_{n,1} & \cdots & s_{n,20} \end{bmatrix}$$

where $s_{i,j}$ is the probability of $j_{th}$ amino acid in $i_{th}$ position of the sequence. DPPI utilizes data augmentation to solve the problem of different length sequences and to increase the diversity of the training samples to improve the robustness of our deep model.

### 7.2.2. Convolutional module

Each convolutional module consists of four layers including convolution layer, rectified linear unit, batch normalization and pooling. All these layers are later explained in this section. In a DPPI model, output of a convolutional module is computed by an expression starting with a convolution layer and ending in a pooling layer:

$$R = \text{Pool}\left(\text{ReLU}\left(\text{Batch}_{\Gamma,\beta}(\text{conv}_M(S))\right)\right),$$

where $R$ is the output vector and $S$ is the input profile.

**Convolution layer.** Let $A$ be a $n \times s$ array of length $n$ with $s$ features in each position. The convolution layer starts by sliding a window of length $m$ along the array to convert $A$ to a $(n - m + 1) \times d$ array $X$ where $d$ is the number of learnable patterns (filters) in the convolution layer. Let $X_{i,k}$ indicate the score of filter $k$, $1 \leq k \leq d$, aligned to position $i$ of array $A$. The tunable parameters, all length $m$, are stored in an $d \times m \times s$ array $M$ where $M_{k,j,l}$ is the coefficient of pattern $k$ at pattern position $j$ and base $l$. Convolution layer computes $X = conv_M(S)$ where

$$X_{i,k} = \sum_{j=1}^{m} \sum_{v=1}^{s} M_{k,j,v} A_{i+j,v}.$$

Specifically, for the first convolution layer the input array is input profile S. In deep learning, column $X_{\cdot,k}$ is known as a 1-dimensional filter scan of filter k using a d-channel input.

**Batch normalization layer.** This layer is applied on the output of the convolution layer and computes a normalized array $Y = \text{Batch}_{\Gamma,\beta}(X)$ where

$$Y_{i,k} = \Gamma_k \widehat{X}_{i,k} + \beta_k.$$

$\Gamma_k$ and $\beta_k$ are parameters of batch normalization for specific feature k and $\widehat{X}_{i,k}$ is the normalized value of $X_{i,k}$ with respect to the mean and variances of feature values in batch $\mathcal{B}$:

$$\widehat{X}_{i,k} = \frac{X_{i,k} - E_{\mathcal{B}}(X_{.,k})}{\sqrt{\text{Var}_{\mathcal{B}}(X_{.,k})}}.$$

$E_{\mathcal{B}}(X_{.,k})$ and $\text{Var}_{\mathcal{B}}(X_{.,k})$ respectively denote the average and variance of feature k in terms of batch $\mathcal{B}$:

$$E_{\mathcal{B}}(X_{.,k}) = \sum_{i=1}^{n-m+1} \sum_{X \in \mathcal{B}} X_{i,k}.$$

Batch normalization layer makes the model insensitive to parameter initialization and speeds up training. The size of array Y is the same as the size of X.

**Rectified Linear Unit.** The rectified linear unit (ReLU) takes the $(n - m - 1) \times d$ output array of batch normalization layer (i.e. Y) and computes an array $Z = \text{ReLU}(Y)$ with the same size as Y. ReLU clamps all the negative values to zero to introduce non-linearity where

$$Z_{i,d} = \max(0, Y_{i,d}).$$

Rectification has been observed to play a major role in the performance of deep learning models in different areas such as image and speech recognition.

**Pooling layer.** The output size of the rectified linear unit, Z, depends on the length of input profile S. A pooling layer is employed to reduce the dimension of Z to a $(n - l_p + 1) \times d$ array R where $l_p$ is the size of pooling window. Array $R = \text{Pool}(Z)$ is computed as the average of all positions $i \leq j \leq i + l_p$ over each feature k where $1 \leq i \leq (n - m + 1) - l_p$:

$$R_{i,k} = \text{Average}(Z_{i,k}, \ldots, Z_{i+l,k}).$$

For PPI prediction, average pooling performs well on its own. It is worth mentioning that the output of the last pooling layer is a $1 \times d$ array.

### 7.2.3. Random projection module

DPPI exploits a Siamese-like convolutional neural network architecture, and contains two identical convolutional modules with the same configuration and parameters for a pair of input profiles S and S'. Let $R_S$ and $R_{S'}$ respectively be the representations learned by convolutional modules for input profiles S and S'.

The role of the random projection module is to take the representation learned by the convolutional module for each of the two proteins and use two different mappings to project theses representations onto two different spaces. The random projection layer (RP) layer consists of two separate fully connected networks. Each fully connected network takes a $1 \times d$ output array R and computes a d-dimensional vector $o = \text{Net}_W(R)$ where:

$$o_i = \sum_{j=1}^{d} W_{j,i} R_{1,j}.$$

$d \times d$ array W is the weight matrix of the layer and vector o is a representation of profile S. Let $W^1$ and $W^2$ respectively denote the weights of first and second network. $d \times d$ array W of RP is split from the middle into two $(d \times d/2)$ arrays $W^{1:1}$ and $W^{1:2}$, and $d \times d$ array $W^2$ is also into two $(d \times d/2)$ arrays $W^{2:1}$ and $W^{2:2}$ where

$$W^{2:1} = W^{1:2}, W^{2:2} = W^{1:1}. \quad (1)$$

Splitting the weights and flipping them helps the model investigate the combination of learned patterns and learn the relationship between proteins with dissimilar and also similar patterns. It also enables DPPI to ignore the order of the input profiles.

The weights $W^1$ are fixed during the training (i.e. untrained), so is $W^2$. Having untrained weights will reduce the number of parameters and thus will speed up the model significantly. It also prevents the Model from over-fitting and thus results in a better classification for the interactions. The output of the random projection layer is a representation for each input protein.

### 7.2.4. Prediction module

Prediction module takes a pair of protein representations $o_S$ and $o_{S'}$, respectively for two protein profiles $S$ and $S'$, and computes a binary score indicating whether two proteins interact. This module first performs an element-wise multiply on two vectors $o_S$ and $o_{S'}$:

$$q = o_S \odot o_{S'},$$

where $\odot$ denotes the element-wise multiplication. Next, a linear layer transforms d-dimensioanl vector $q$ into a binary output score $f(S, S') = \text{Linear}_w(q)$. A linear layer is a $(1 + d)$-dimensional vector of tunable weights $w$, where the output score is calculated as the linear combination

$$f(S, S') = w_0 + \sum_{k=1}^{d} w_k q_k,$$

where weight $w_k$ is the weight of $q_k$'s contribution to the output, and $w_0$ is an additive bias term.

**Remark.** As we mentioned in previous section, DPPI switch the random weights in RP module to help DPPI overcome several fundamental limitations, including insensitivity to the order of input profiles and encoding patterns for low-similar proteins. Specifically, if $q$ and $q'$ denote the output of element-wise multiplication respectively for input profiles $(S, S')$ and $(S', S)$, we have:

$$\begin{aligned} q &= o_S \odot o_{S'} = (R_S \times W^{1:1}. R_S \times W^{1:2}) \odot (R_{S'} \times W^{2:1}. R_{S'} \times W^{2:2}) \\ q' &= o_{S'} \odot o_S = (R_{S'} \times W^{1:1}. R_{S'} \times W^{1:2}) \odot (R_S \times W^{2:1}. R_S \times W^{2:2}) \\ &= (R_{S'} \times W^{2:2}. R_{S'} \times W^{2:1}) \odot (R_S \times W^{1:2}. R_S \times W^{1:1}) \quad \text{w.r.t} \quad (1) \\ &\Rightarrow q = q'. \end{aligned}$$

Above equation concludes that linear layer would have the same input vector regardless of the order of input profiles.

Furthermore, switching the weighs let the model to learn patterns for interacting proteins those who have different patterns of amino acids (e.g. heterodimeric proteins). Let $W_{d,k}^{1}$ and $W_{d,k}^{2}$ denote the (d×1) weight vector for filter k, $1 \leq k \leq d/2$, respectively in first and second network of the RP module. By doing an element-wise multiplication in prediction module the value for filter k is computed as:

$$q_k = (R_S \times W_{d,k}^{1:1}).(R_{S'} \times W_{d,k}^{2:1}) \qquad (2)$$

Because $W_{d,k}^{1:1} \neq W_{d,k}^{2:1}$, even having a low value for one of the representation arrays $R_S$ and $R_{S'}$ will not prevent having a large value for $q_k$, which is beneficial when interacting proteins have a different composition of amino acids. That is DPPI can detect the combinations of different patterns between interacting proteins rather than just being forced to detect similar patterns.

## 7.3. Training pipeline

We trained DPPI described in Figure 7.1 using error backpropagation with stochastic gradient descent algorithm with momentum. Our implementation uses the software package Torch7/Lua [99].

### 7.3.1. Data augmentation

Allowing the input sequences to have varying lengths creates several complications in designing a multi-layer architecture and using pooling and batch normalization. To avoid these issues, we crop each protein sequence into overlapping subsequences of length 512 with an overlapping margin of 256. Specifically, the first subsequence starts at position 1, the second one starts at position 256 and so on until the whole sequence is covered. Therefore, each subsequence has overlap of length 256 with the previous subsequence and the next one. Thus, for a sequence profile with length $l_s$, cropping generates C different subsequences where

$$C = \begin{cases} 1 & l_s \leq 512 \\ \lceil l_s/256 \rceil - 1 & otherwise \end{cases}.$$

These data augmentation method is observed to improve the generalization of the convolutional layers and prevent the overfitting in image processing [100].

During the training, each of the subsequences are treated as a separate protein and therefore for a pair of interacting proteins $S$ and $S'$, we train the model to be able to predict their interaction by given any pair of their corresponding crops. During the evaluation, however, we are interested in a single binary value $f(S, S')$. Therefore, we compute $f(S, S')$ to be the maximum predicted score over all crop combinations of these two proteins $S$ and $S'$.

### 7.3.2. Back-propagation stages

DPPI is trained by a gradient-based backpropagation method to minimize the prediction error with respect to a loss function. After computing the $f(S, S')$, the loss value is back propagated through the layers and used to calculate the gradient of the loss function with respect to the parameters. This gradient is used to update the parameters values. For more details on backpropagation algorithm reader is referred to [101, 102].

Let θ denotes the vector of all parameters including filters $M$, parameters $\Gamma$ and $\beta$ and network weights $w$. DPPI updates $\theta$ with respect to the objective function using stochastic gradient descent (SGD):

$$\theta^{t+1} = \theta^t - \eta \nabla f(S, S'),$$

where $\theta^{t+1}$ indicates the updated version of current parameters $\theta^t$ and $\eta$ is learning rate parameter.

### 7.3.3. Training objective

Suppose we are given N training pairs $((S, S')^1, y^1), \ldots, ((S, S')^N, y^N)$ where $(S, S')^i$ is a pair of input protein profiles and $y^i = y((S, S')^i)$ is the binary training label where:

$$\begin{cases} y^i = 1 & S \text{ and } S' \text{ interact} \\ y^i = 0 & \text{otherwise} \end{cases}.$$

Let $f^i$ indicate the corresponding DPPI prediction $f^i = f((S, S')^i)$ with respect to $\theta$. For training our model, we aim to approximately minimize the following regularized objective function:

$$\sum_{i=1}^{N} LOSS(f^i, y^i) + \lambda \|\theta\|_2,$$

where $\theta$ is the vector of all parameters, $\lambda$ is the regularization parameter which controls the model complexity. $\|.\|_2$ denotes the *L*2. *L*2-norm regularization prevents the over-fitting by reducing the parameter space. The loss function is defined using a logistic regression function over all protein pair scores:

$$\sum_{i=1}^{N} \ln\left(1 + e^{-(f^i \times y^i)}\right).$$

## 7.4. Results

To evaluate DPPI's ability to predict direct physical protein interactions in an unbiased manner, we obtained high-quality PPI benchmark data from human and yeast [18].

### 7.4.1. Comparison with other methods

Here, we trained our model on a pre-defined set of binary labeled positive and negative protein interactions, and then tested predictive performance on a held-out set of protein interactions. To remove sequence homology bias, we focused our attention to the most stringent benchmark where neither protein in a test pair was found with greater than 40% sequence identity in the training data (referred to as C3[18, 103]). We performed 10-fold cross-validation for each species and demonstrate that DPPI outperforms all other sequence-based methods (Figure 7.2-A).

Next, we sought to evaluate if DPPI improves with additional training examples. We curated a large dataset consisting of direct physical PPIs from 11 different species[67]. We randomly sampled 1k, 2.5k, 5k, 10k, 25k and 50k positive interactions for six different training sizes, and then assembled test data using 25% of the training size from the remaining interactions. We reduced sequence homology bias by removing any protein pair in the test data with greater than 40% sequence identity to the training data. Similar to Hamp and Rost[18], for each positive interaction, 10 negatives were generated by randomly sampling from all proteins (10x random negative). We found that DPPI continues to learn and improve mean average precision with the addition of more examples (Figure 7.2-B). Furthermore, in contrast to the leading kernel based method, DPPI scales efficiently with increasing numbers of training examples (Figure 7.2-C).

### 7.4.2. Evaluating binding affinities

To assess real-world applications of DPPI, we considered three different but related tasks. First, we wondered if DPPI can be used for "deorphanization" tasks or matching receptors to their ligands. We sought to computationally predict the receptor for IL34, which was discovered using an exhaustive genome-wide screen involving the cloning and expression of a large number of receptors that ultimately identified CSF1R104. We trained DPPI using all known direct physical human PPIs from multiple sources67, 105, resulting in 211,744 positive interactions and 5x random negative. We removed any protein pair with more than 40% sequence similarity to IL34 from the training data, and tested the interaction between IL34 and each of the 135 receptors obtained from the KEGG Cytokine-Receptor Pathway. We found that CSF1R ranked 2nd in our predictions, with a percentile score of 0.993. This suggests that DPPI may provide a valuable means for discovering and prioritizing protein interactions for low-throughput experimental studies.

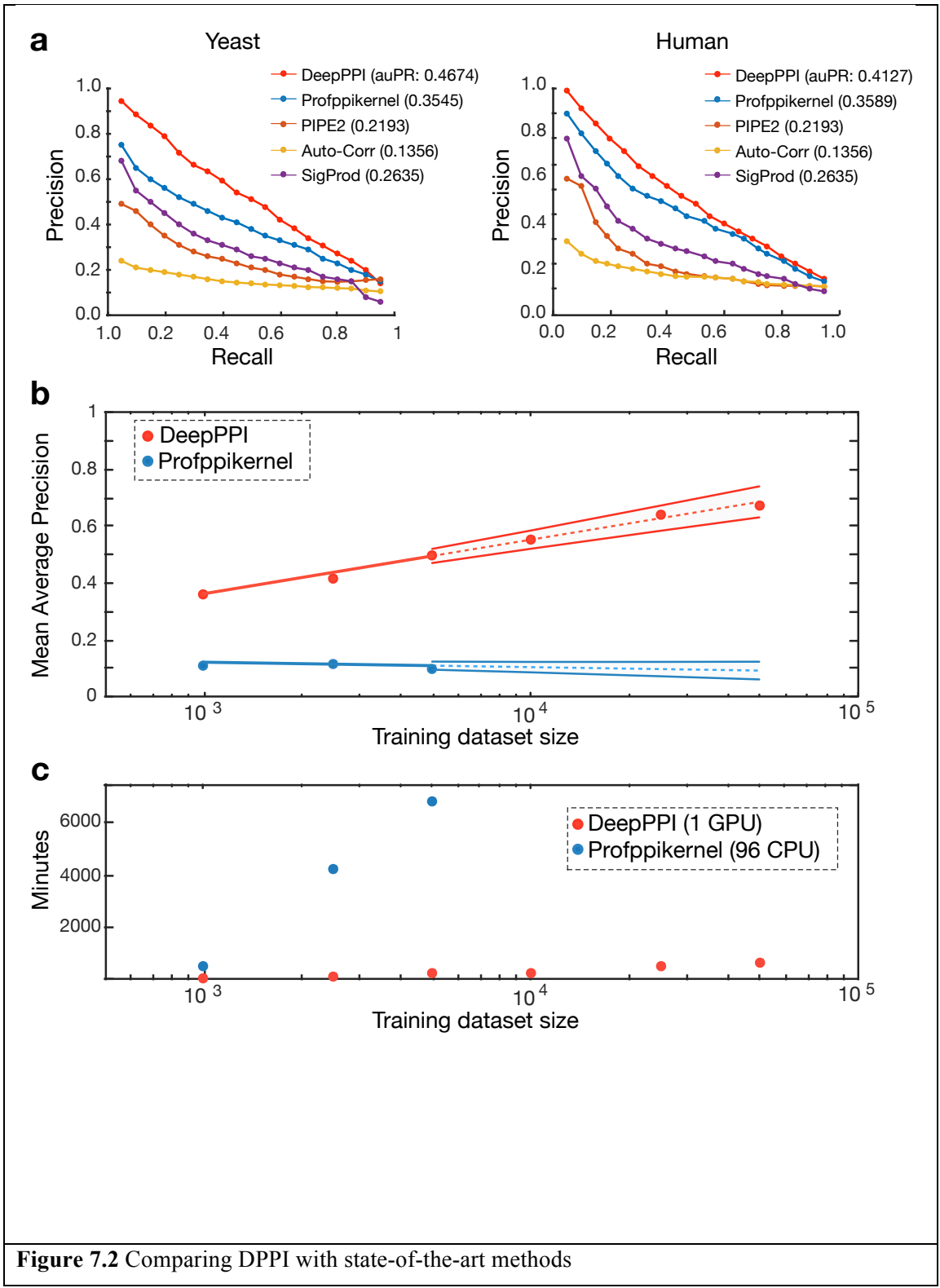

**Figure 7.2** Comparing DPPI with state-of-the-art methods

Second, motivated by our initial result, we sought to characterize the relationship between percentile scores and binding affinities. Cytokines, such as IL-2[106] and IL-13[107], have been engineered to bind their cognate receptors at different binding affinities. We applied DPPI to engineered variants of IL-2 and IL-13, and their cognate receptors. To compare results between cytokines, we simulated 1000 sequences by randomly permuting the amino acid positions engineered to confer the strongest binding affinity for each cytokine. We found a significant correlation between the percentile scores and the experimentally determined binding affinities (Figure 7.3-A). These results suggest that DPPI may help to identify specific amino acid substitutions resulting in increased or decreased binding affinities.

Finally, we sought to directly test if DPPI can function as a generalizable framework to model protein-protein interaction binding affinities. We analyzed results from a recent deep mutational scanning study, where the authors extensively made single amino acid substitutions to a computationally designed protein inhibitor of the 1918 H1N1 influenza hemagglutinin (HA)[108]. The authors report enrichment values that are proportional to the change in binding free energy due to the substitutions. We trained DPPI with one half of the variants, and then calculated the percentile scores of the interactions between HA and the remaining inhibitor variants. We found that the predicted percentile scores where significantly correlated with the enrichment values of the inhibitor (Figure 7.3-B). Thus, even without using structure information DPPI can effectively model protein-protein interaction binding affinities.

Taken together, our deep learning framework serves as a principled computational approach to model and predict protein interactions from sequence and is generalizable to numerous applications.

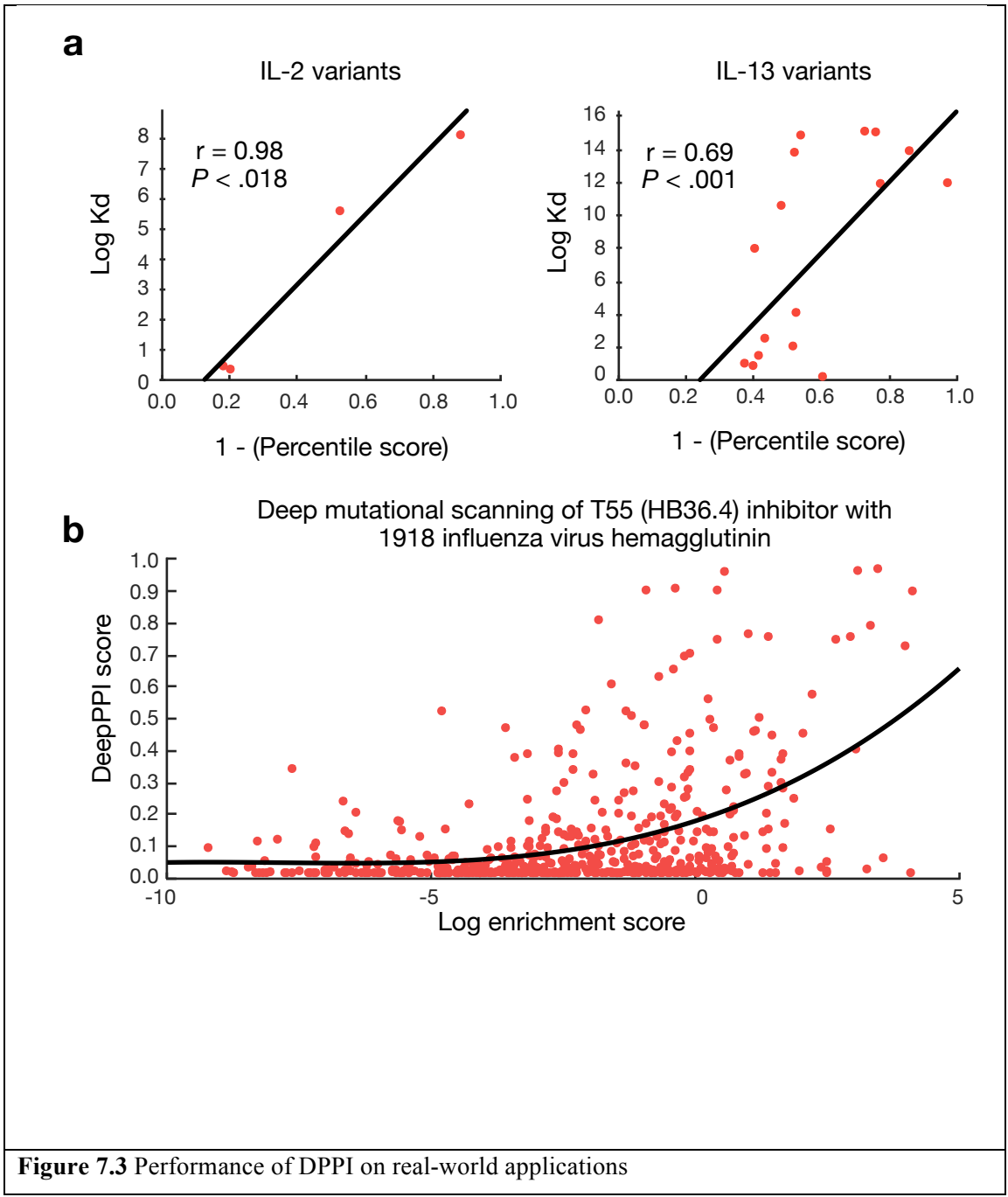

**Figure 7.3** Performance of DPPI on real-world applications